\documentclass[fleqn,usenatbib]{mnras}

\usepackage{newtxtext,newtxmath}

\usepackage[T1]{fontenc}

\DeclareRobustCommand{\VAN}[3]{#2}
\let\VANthebibliography\thebibliography
\def\thebibliography{\DeclareRobustCommand{\VAN}[3]{##3}\VANthebibliography}

\usepackage{graphicx}	
\usepackage{amsmath}	
\usepackage[detect-all]{siunitx}
\usepackage{bm}
\usepackage{booktabs}
\usepackage{dblfloatfix}
\usepackage{siunitx}
\usepackage{dcolumn}
\usepackage[flushleft]{threeparttable}

\newcommand{\nStars}{194,879}
\newcommand{\nELL}{369}
\newcommand{\nECL}{15}
\newcommand{\nSH}{356}

\newcommand{\nGaiaR}{313}
\newcommand{\Pmin}{0.25}
\newcommand{\Pmax}{143.19}
\newcommand{\Pmed}{9.3}
\newcommand{\nAPOGEE}{13}
\newcommand{\nAPOGEEmultiple}{9}
\newcommand{\nLAMOST}{57}
\newcommand{\nLAMOSTmultiple}{12}
\newcommand{\nRAVE}{13}
\newcommand{\nameRVbad}{ASASSN-V J080826.59-055109.9, J084328.53+402247.5, J061635.56+231909.4, and J161803.43+420416.8}
\newcommand{\nRVbad}{4}
\newcommand{\eclPmed}{11.86}

\newcommand{\McminMed}{0.2}
\newcommand{\nMCgtONE}{14}
\newcommand{\SHinvalidR}{14}
\newcommand{\nXray}{9}

\newcommand{\Pclass}{$P_{\rm{class}}$}
\newcommand{\chiELL}{$\chi^2_{\rm{ell}}$}
\newcommand{\chiCOS}{$\chi^2_{\rm{cos}}$}
\newcommand{\Ratio}{$\chi^2_{\rm{ell}}/\chi^2_{\rm{cos}}$}
\newcommand{\chiLIN}{$\chi^2_{\rm{line}}$}
\newcommand{\chiMED}{$\chi^2_{\rm{med}}$}
\newcommand{\RatioLT}{$\chi^2_{\rm{line}}/\chi^2_{\rm{med}}$}
\newcolumntype{d}[1]{D{.}{.}{#1}}

\title[ASAS-SN Ellipsoidal Variables]{High Tide: A Systematic Search for Ellipsoidal Variables in ASAS-SN}

\author[D. M. Rowan et al.]{D. M. Rowan$^{1,2}$\thanks{E-mail: rowan.90@osu.edu},
K. Z. Stanek$^{1,2}$,
T. Jayasinghe$^{1,2}$,
C. S. Kochanek$^{1,2}$,
Todd A. Thompson$^{1,2}$,
\newauthor B. J. Shappee$^{3}$,
T. W. -S. Holoien$^{4}$,
J. L. Prieto$^{5,6}$
\\
$^{1}$Department of Astronomy, The Ohio State University, 140 West 18th Avenue, Columbus, OH, 43210, USA\\
$^{2}$Center for Cosmology and Astroparticle Physics, The Ohio State University, 191 W. Woodruff Avenue, Columbus, OH, 43210, USA\\
$^{3}$Institute for Astronomy, University of Hawaii, 2680 Woodlawn Drive, Honolulu, HI 96822, USA\\
$^{4}$Carnegie Observatories, 813 Santa Barbara Street, Pasadena, CA 91101, USA\\
$^{5}$N\'ucleo de Astronom\'ia de la Facultad de Ingenier\'ia y Ciencias, Universidad Diego Portales, Av. Ej\'ercito 441, Santiago, Chile\\
$^{6}$Millennium Institute of Astrophysics, Santiago, Chile\\
}

\date{Accepted XXX. Received YYY; in original form ZZZ}

\pubyear{2021}

\begin{document}
\label{firstpage}
\pagerange{\pageref{firstpage}--\pageref{lastpage}}
\maketitle

\begin{abstract}

The majority of non-merging stellar mass black holes are discovered by observing high energy emission from accretion processes. Here we pursue the large, but still mostly unstudied population of non-interacting black holes and neutron stars by searching for the tidally-induced ellipsoidal variability of their stellar companions. We start from a sample of about 200,000 rotational variables, semi-regular variables, and eclipsing binary stars from the All-Sky Automated Survey for Supernovae (ASAS-SN). We use a $\chi^2$ ratio test followed by visual inspection to identify \nELL{} candidates for ellipsoidal variability. We also discuss how to combine the amplitude of the variability with mass and radius estimates for observed stars to calculate a minimum companion mass, identifying the most promising candidates for high mass companions. 

\end{abstract}




\section{Introduction}

There are an estimated $10^8$--$10^9$ stellar mass black holes (BHs) in the Milky Way \citep{Timmes96, Wiktorowicz19}. Stellar mass black holes are thought to be produced largely by direct collapse without a supernova \citep[e.g., ][]{Ugliano12, Pejcha15}. Two candidates for such a failed supernova were identified by \citet{Gerke15} and \citet[also see \citealp{Adams17, Basinger20}]{Neustdat21}, as a part of the search for ``vanishing'' stars proposed by \citet{Kochanek08}. Double neutron star (NS) mergers, such as G170817 \citep{Abbott17GW170817} are also a channel of stellar mass BH formation \citep{Pooley18}. Searches for stellar mass BHs are crucial for characterizing the end states of evolved massive stars, the underlying compact object mass distribution, and the intermediate evolutionary states before compact object mergers.

Stellar mass black holes are typically observed in X-ray binary systems \citep[e.g., ][]{Remillard06} or in gravitational wave detections of compact object mergers \citep[e.g.][]{Abbott16}. The \texttt{BlackCat} catalog currently includes 69 X-ray emitting BH binaries \citep{Corral16}. Most of those systems with mass estimates have masses of 5--$10M_{\odot}$ with a low mass gap from $\sim$ 2 -- 5$ M_{\odot}$ that separates the neutron star and stellar mass black hole populations \citep{Bailyn98, Ozel10, Farr11, Kochanek14}. Between the first Gravitational Wave Source Catalog \citep[GWTC-1,][]{Abbott19} and the expanded catalog \citep[GWTC-2,][]{Abbott20}, 50 gravitational wave merger events have been detected and the masses of merger components also suggest a mass gap between BHs and NSs \citep{Fishbach20, LIGOCollab20}.

Interacting binaries and gravitational wave sources are, however, very biased probes for constructing a census of compact objects. The conditions leading to mass transfer or mergers are rare, so such systems represent a very small fraction of the total BH binary population \citep{Tanaka02, Wiktorowicz19}. Even for systems with X-ray emission, there are extended periods of quiescence \citep[e.g., GRS 1915$+$105,][]{Huppenkothen17, Tucker18}. Understanding interacting and merging binaries requires identifying and understanding the population of non-interacting binary black holes.

However, the detection of non-interacting black holes is inherently challenging due to their electromagnetically dark nature. There are estimates that $10^2$--$10^5$ long-period BH binaries will be detectable with Gaia astrometry \citep{Breivik17, Mashian17, Yalinewich18, Yamaguchi18}. Short-period, detached BHs with non-compact object companions may be detected in TESS photometry through a combination of microlensing events, ellipsoidal variations, and relativistic beaming \citep{Masuda19, Gomel20}. Microlensing surveys can also be used to identify compact object candidates \citep[see ][]{Paczynski86, Gould14, Abdurraham21}. 

Because dynamical processes in dense stellar systems can drive the formation of BH binaries, globular clusters should have more BH binaries than the field per unit stellar mass \citep{Askar18}. \citet{Giesers18} identified a BH with $M=4.36\pm0.41\ M_{\odot}$ in the globular cluster NGC 3201 and two additional BHs in NGC 3201 were reported in \citet{Giesers19}. 

Several candidate non-interacting BHs in the field have been reported and debated in the last few years. \citet{Liu19} detected radial velocity variations in a B star, LB-1, suggesting the presence of a $68^{+11}_{-13}~M_{\odot}$ BH companion. Subsequent analysis has suggested a lower companion mass 2--3$~M_{\odot}$ BH, but the nature of the system remains unclear \citep{ElBadry20, Irrgang20, Shenar20, Lennon21}. \citet{Rivinius20} claimed that HR 6819 is a hierarchical triple containing stellar mass black hole, but the system was later suggested to be a binary of a slowly rotating B and a rapidly rotating Be star \citep{ElBadry21}. 

Other systems are still best-described with non-interacting BH companions. \citet{Thompson19} reported a stellar mass BH with $M=3.3^{+2.8}_{-0.7}\ M_{\odot}$ orbiting the spotted red giant 2MASS J05215658$+$4359220. They identified this system by searching sparse APOGEE radial velocity (RV) measurements for systems with high binary mass function and photometric variability in ASAS-SN. Most recently, \citet{Jayasinghe21} combined archival spectroscopic orbits with photometric observations of ellipsoidal variability to identify a $3.04\pm0.06~M_\odot$ black hole orbiting the nearby, bright red giant V723~Mon. In both cases, spectroscopic surveys were the first step of the compact object search. 

Another approach to search for compact objects in binary systems is to search for ellipsoidal variables (ELLs) in time-domain photometric surveys. Ellipsoidal variability occurs due to the tidal distortions of a star by its binary companion. ELL light curves have a characteristic double-peaked structure with typically uneven minima. Figure \ref{fig:unicorn} shows the KELT \citep{Pepper07} light curve of V723~Mon \citep{Jayasinghe21} as an example. While RV measurements are needed to confirm the nature of the variability \citep{Soszynski04}, a photometric search for ELLs offers a practical starting point to study the Milky Way stellar mass BH population. 

Here we search for ELLs using the All-Sky Automated Survey for Supernovae \citep[ASAS-SN,][]{Shappee14,Kochanek17, Jayasinghe18}. In Section \S\ref{sec:observations} we describe the initial selection of targets from the ASAS-SN variable stars catalog. We use the analytical model of \citet{Morris93} to search for ELL candidates in Section \S\ref{sec:search_method}. ELL light curves can closely resemble those of eclipsing binaries, spotted variables, and RV Tauri stars. In Section \S\ref{sec:visual_inspection} we describe how we visually inspect the initial candidates to produce our final catalog. In Section \S\ref{sec:ELL+ECL} we describe \nECL{} systems that show both ellipsoidal variations and eclipses. We combine the photometric properties of the observed stars from the \citet{Anders19} \texttt{StarHorse} catalog with our ELL model amplitudes to derive minimum companion masses in Section \S\ref{sec:minimum_companion_mass}. We cross-match our ELL catalog with radial velocity catalogs and X-ray catalogs in Section \S\ref{sec:rv_xray}. Finally, we present a summary of our results in Section~\S\ref{sec:summary}. 

\section{Searching for Ellipsoidal Variables} \label{sec:search_method}

The ASAS-SN $V$-band observations made between 2012 and mid 2018 have been used to classify $\sim 426,000$ variable stars, including $\sim 219,000$ new discoveries \citep{Jayasinghe21IX}. Since the number of ELL variables is expected to be small relative to other types of variability, ELLs were not included in the random forest classification used by \citet{Jayasinghe19}. Some ELLs were visually identified as a part of \citet{Pawlak19}, but most will have been classified as eclipsing binaries or rotational variables. We use an analytic model to search for ELL light curves in ASAS-SN and validate the candidates with visual inspection. In the absence of RV measurements, we combine the ELL model with photometric estimates of the stellar properties to derive a minimum companion mass for the ELL candidates.

\subsection{ASAS-SN Search Catalog} \label{sec:observations}

We begin with the ASAS-SN catalog of variable stars \citep{Jayasinghe18}. Since ELLs can be confused with other variable classifications, we make a broad selection from the catalog in classification probability \Pclass{} and period $P$. We selected the following stars for our search:
\begin{itemize}
    \item W UMa (EW) binaries with \Pclass{}$> 0.4$
    \item Beta Lyrae (EB) binaries with \Pclass{}$> 0.4$
    \item Rotational variables (ROT) with \Pclass{}$>0.4$
    \item Semi-regular variables (SR) with \Pclass{}$>0.4$ and $P<60$ days
    \item Ellipsoidal variables (ELL) with \Pclass{} $>0.9$ and amplitude $<0.4$
\end{itemize}
\noindent The $P<60$ day cutoff for SRs is chosen to reduce contamination from dust-producing pulsating AGB and RGB stars \citep{Alard01, McDonald18, Jayasinghe21IX}. This led to an initial search catalog of \nStars{} stars. We use the periods from \citet{Jayasinghe19} to phase-fold the $V$-band light curves.

\subsection{Analytical Model for Ellipsoidal Modulations}
\begin{figure}
    \centering
    \includegraphics[width=\linewidth]{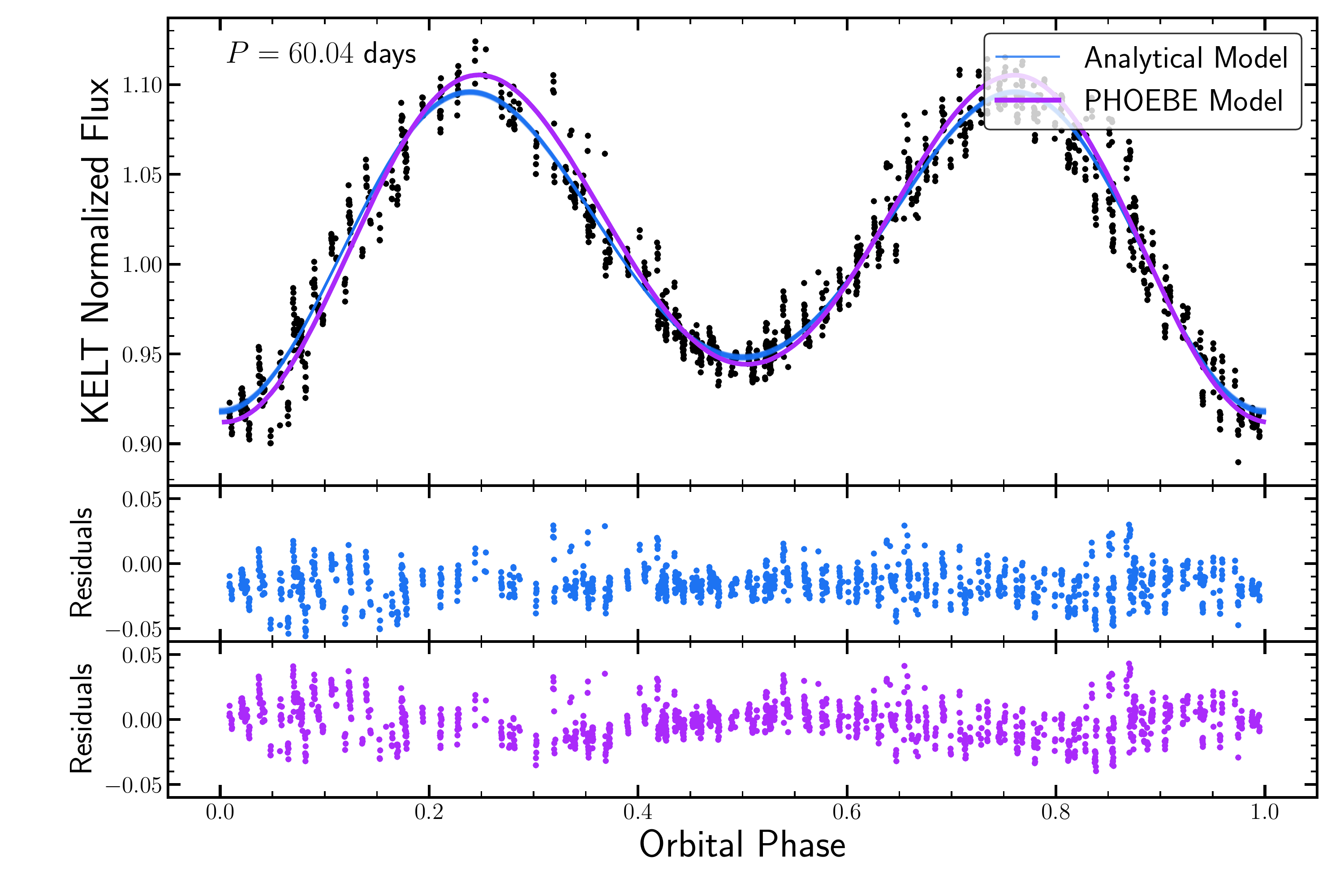}
    \caption{KELT light curve for V723~Mon \citep{Jayasinghe21}. The analytical model used in Section \S\ref{sec:search_method} (blue), is consistent with the PHOEBE model fit by \protect{\citet{Jayasinghe21}} (purple). The middle panel shows the residuals of the analytical model and the bottom panel shows the residuals of the PHOEBE model.}
    \label{fig:unicorn}
\end{figure}
\begin{figure*}
    \centering
    \includegraphics[width=\linewidth]{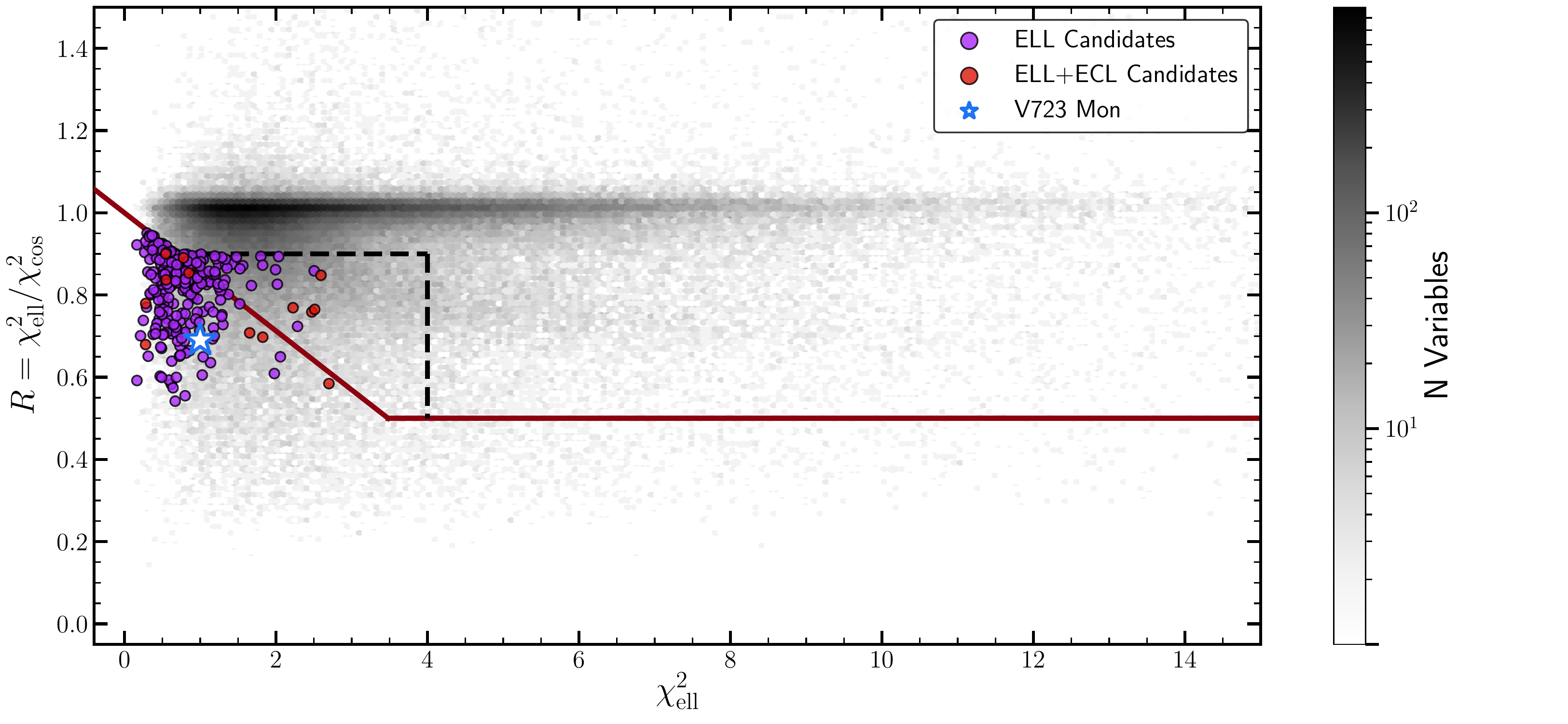}
    \caption{We compare the \chiELL{} of the fit to the analytical model of ellipsoidal variability to the \chiCOS{} of a simple cosine fit. The red line indicates our visual inspection sample corresponding to $\sim 5\%$ of the total sample. The V723~Mon KELT light curve \protect{\citep{Jayasinghe21}} is shown as a star and would be selected by this cut. Points below the red line are selected for visual inspection. Purple and red points are the ELL and ELL+ECL systems remaining after visual inspection. We also inspect stars in the region bounded by the black dotted lines to assess the effectiveness of the original cut. We find the majority of ELL candidates are selected by our original $\chi^2$ criterion.}
    \label{fig:chi_compare}
\end{figure*}
We fit each light curve with a series of analytic models to identify the best ELL candidates. Ellipsoidal modulations have a characteristic double-peaked structure with uneven maxima where the fractional luminosity changes can be represented by a discrete Fourier series. The first three terms are
\begin{equation} \label{eqn:analytical model}
\begin{split}
    \frac{\Delta L}{\bar{L}} = &\frac{\alpha_1}{\bar{L}/L_0} \left(\frac{R_*}{a}\right)^4 q \left(4\sin i-5\sin^3i\right) \cos \phi\\
    &- \frac{1}{\bar{L}/L_0} \bigg[\alpha_2 \left(\frac{R_*}{a}\right)^3 q\sin^2i\\
    &+\beta_2 \left(\frac{R_*}{a}\right)^5 q\left( 6\sin^2i-7\sin^4i\right)\bigg]\cos2\phi\\
    &-\frac{5}{3}\frac{\alpha_1}{\bar{L}/L_0} \left(\frac{R_*}{a} \right)^4 q\sin^3i\cos3\phi
\end{split}
\end{equation}
where $R_*$ is the radius of the luminous primary, $a$ is the binary semi-major axis, $q=M_2/M_1$ is the mass ratio of the secondary to the photometric primary, and $i$ is the inclination \citep{Morris93, Gomel20}. The average luminosity $\bar{L}$ is given by
\begin{equation}
    \bar{L} = L_0 \left( 1 + \frac{1}{9}\alpha_2\left(\frac{R_*}{a}\right)^3(2+5q)\left(2-3\sin^2i\right)\right)
\end{equation}
and the coefficients
\begin{equation}
\begin{split}
    \alpha_1 &= \frac{15u(2+\tau)}{32(3-u)},\\
    \alpha_2 &= \frac{3(15+u)(1+\tau)}{20(3-u)}, \rm{and}\\
    \beta_2 &= \frac{15(1-u)(3+\tau)}{64(3-u)}
\end{split}
\end{equation}
depend on the linear-limb darkening coefficient $u$ and the gravity darkening coefficient $\tau$ \citep{Gomel20}. Both $u$ and $\tau$ depend on the temperature, surface gravity, and composition of the star, where we use the tables from \citet{Claret11}. 

We use an approximate analytic model rather than detailed light curve modeling tools \citep[e.g., PHOEBE, ][]{Conroy20} for speed. Figure \ref{fig:unicorn} compares a Markov Chain Monte Carlo (MCMC) fit of the analytic model as compared to the PHOEBE light curve model of the KELT light curve of V723~Mon in \citet{Jayasinghe21}. The analytical model reproduces the PHOEBE model fit with sufficient accuracy for a fraction of the computational cost.

We first fit the analytical model in terms of the amplitudes of the three Fourier terms rather than the physical properties. Since the period calculated by \citet{Jayasinghe19} may correspond to $P/2$ we fit the analytical model to light curves folded at both $P$ and $2P$, and use the relative goodness of fit to determine which corresponds to the ELL period. We compare the \chiELL{} of the ELL fit to a cosine fit with \chiCOS{} and compute the ratio $R=$\Ratio{}. ELLs with uneven minima will have a low \chiELL{} and $R<1$. Figure \ref{fig:chi_compare} shows $R$ as a function of \chiELL{} for all variables in our search catalog. Since we are interested in finding a relatively ``clean'' sample of ELLs and not a complete catalog of ELLs in ASAS-SN, we first make a cut, indicated by the red line, that encompasses $\sim 5\%$ of the total sample. We visually inspect all light curves for variables that fall below this line. 

\begin{figure}
    \centering
    \includegraphics[width=\linewidth]{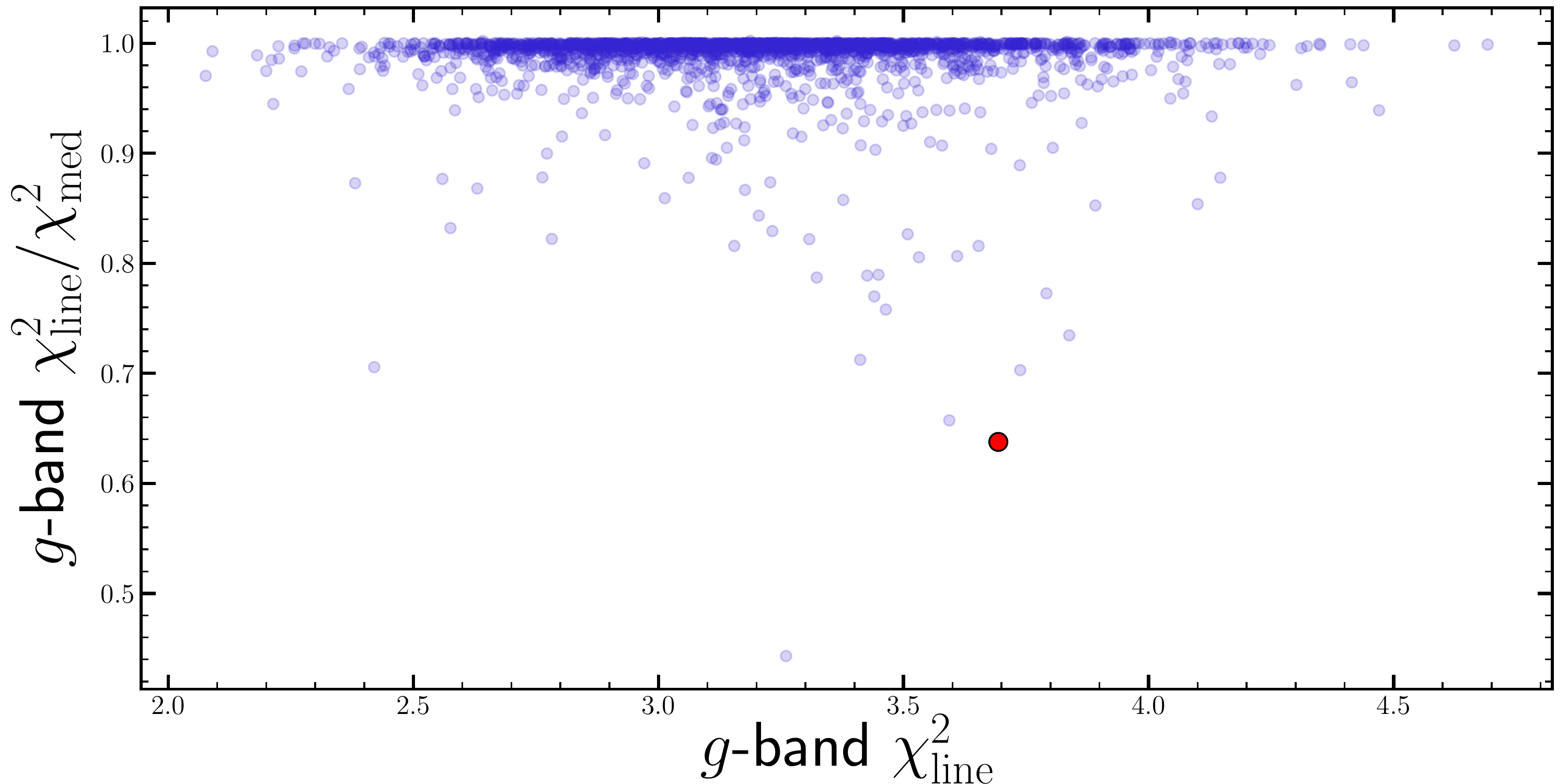}
    \caption{We use the $\chi^2$ ratio of a linear fit \chiLIN{} to a constant \chiMED{} to identify spotted stars with long-term variability. Blue points correspond to all stars selected from the ELL $\chi^2$ ratio cut after visual inspection of the V-band data.The red point corresponds to ASASSN-V J021306.19-324658.8, a rotational variable identified during visual inspection and shown in Figure \ref{fig:rot_example}.}
    \label{fig:rot_chi2}
\end{figure}

We used two additional simple empirical cuts to help eliminate spotted variables. These stars often show drifts in their mean magnitude as the spots evolve. We computed a \chiMED{} of the light curve about the median magnitude and a \chiLIN{} of a light curve about a linear fit in time. We expect spotted stars to have a ratio $R_{\rm{LT}}=$\RatioLT{}$<1$. Figure \ref{fig:rot_chi2} shows the distribution of stars in the $g$-band \chiLIN{} and the ratio \chiLIN{}$/$\chiMED{}. Stars with the ratio significantly below unity tend to be spotted stars, as illustrated by the example marked in Figure \ref{fig:rot_chi2} and shown in Figure \ref{fig:rot_example}.

\subsection{Light Curve Visual Inspection} \label{sec:visual_inspection}

While the $\chi^2$ search method is effective at identifying ELLs with uneven minima, the majority of sources are still non-ELLs, with eclipsing binary, rotational variable, and long-period pulsator light curves significantly contaminating our sample. For the visual inspection, we supplement the $V$ band light curves with ASAS-SN $g$-band light curves and TESS light curves from either the SPOC \citep{Caldwell20} or QLP \citep{Huang20a, Huang20b} reduction pipelines. Before visual inspection we phase all light curves such that the photometric minimum occurs at $\phi=0$. 

\begin{figure}
    \centering
    \includegraphics[width=\linewidth]{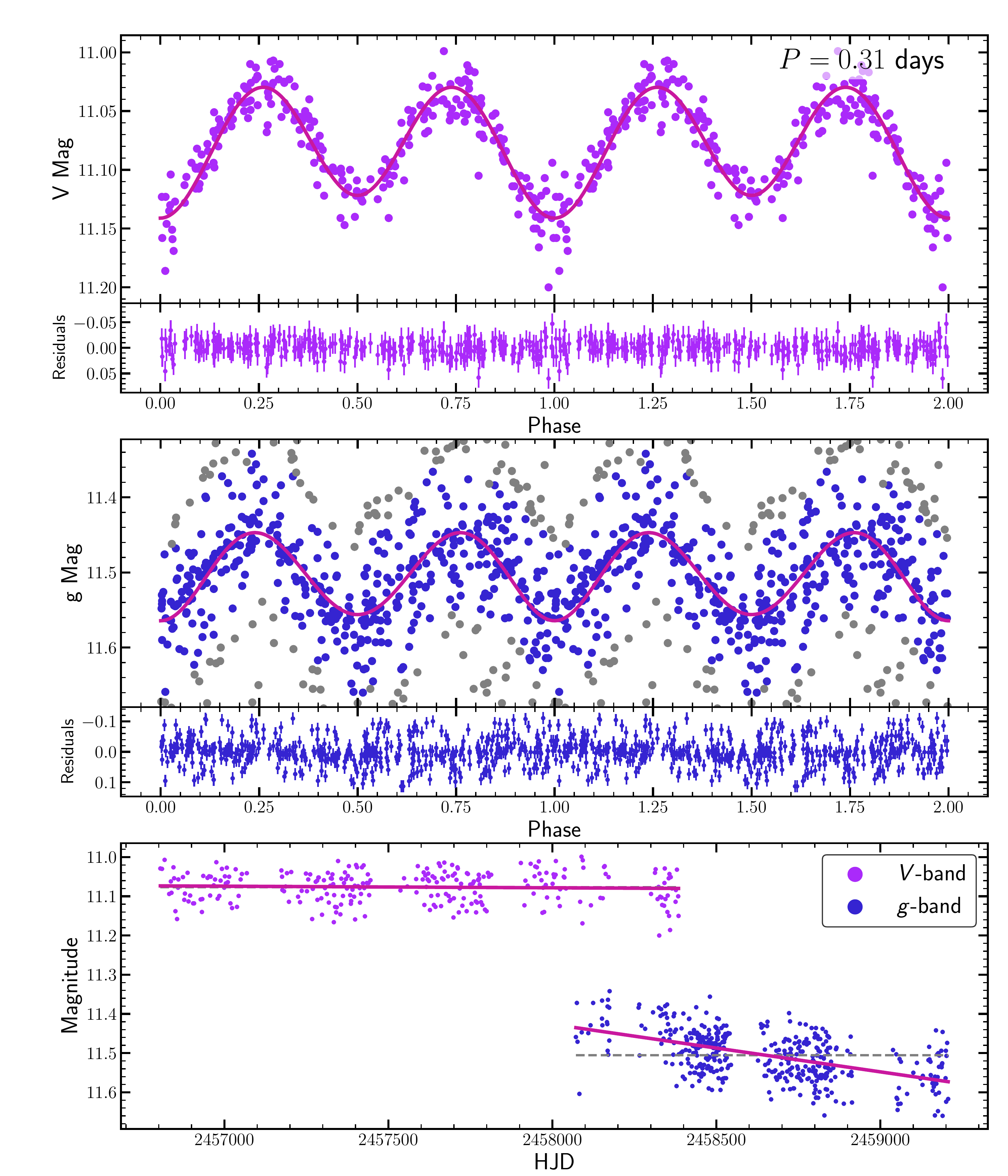}
    \caption{ASAS-SN $V$-band (top), $g$-band (middle), and unfolded (bottom) light curves for ASASSN-V J021306.19-324658.8. Solid lines in phase-folded light curves correspond to the ELL model. Gray points are points clipped at 5$\sigma$ from the binned light curve. The bottom panels shows the linear fits in time for each band (magenta line) compared to the median magnitude (gray line). The $V$-band light curve was selected as part of our visual inspection sample but is likely a spotted star because of the long-term evolution shown in average $g$-band flux.}
    \label{fig:rot_example}
\end{figure}

Spotted stars can have a double-peaked light curve with two uneven minima.  Many also show long term trends as illustrated by the variable in Figure \ref{fig:rot_example}. The $g$-band light curves typically show more scatter and longer linear trends. This is likely due to the contribution of the calcium H and K lines created by chromospheric activity to the $g$-band.

\begin{figure}
    \centering
    \includegraphics[width=\linewidth]{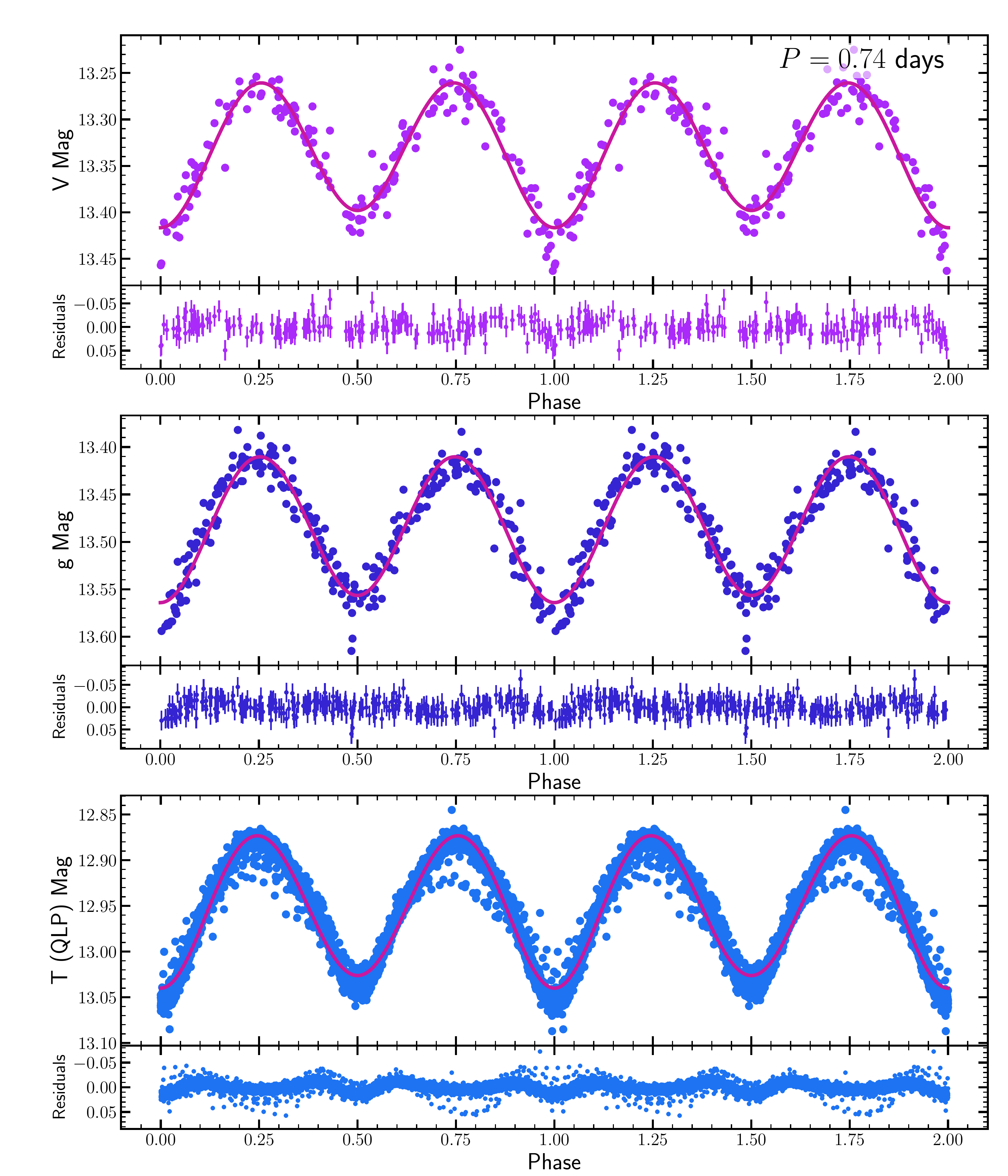}
    \caption{ASAS-SN $V$-band (top), $g$-band (middle), and TESS QLP (bottom) light curves for ASASSN-V J192943.61+641153.4. Solid lines in the light curves correspond to least squares ELL fit. The smaller panels below each light curve show the residuals. This light curve is more consistent with an eclipsing binary than an ELL variable.}
    \label{fig:eb_example}
\end{figure}

Eclipsing binaries can also show two uneven minima separated by 0.5 in phase. Figure \ref{fig:eb_example} shows an example EW light curve, ASASSN-V J192943.61+641153.4, that was selected by our $\chi^2$ search. In this case, the TESS light curve closely resembles known contact binaries such as YY Eri \citep[Figure 3, ][]{Maceroni82}.  When TESS data is unavailable, the residuals in the $V-$ and $g$-band light curves near the photometric minimum are effective at discriminating ELL from EB/EW. Comparing the three bands, the residuals of the ELL fit to ASASSN-V J192943.61$+$641153.4 show correlated residuals near $\phi=0$.  

\begin{figure}
    \centering
    \includegraphics[width=\linewidth]{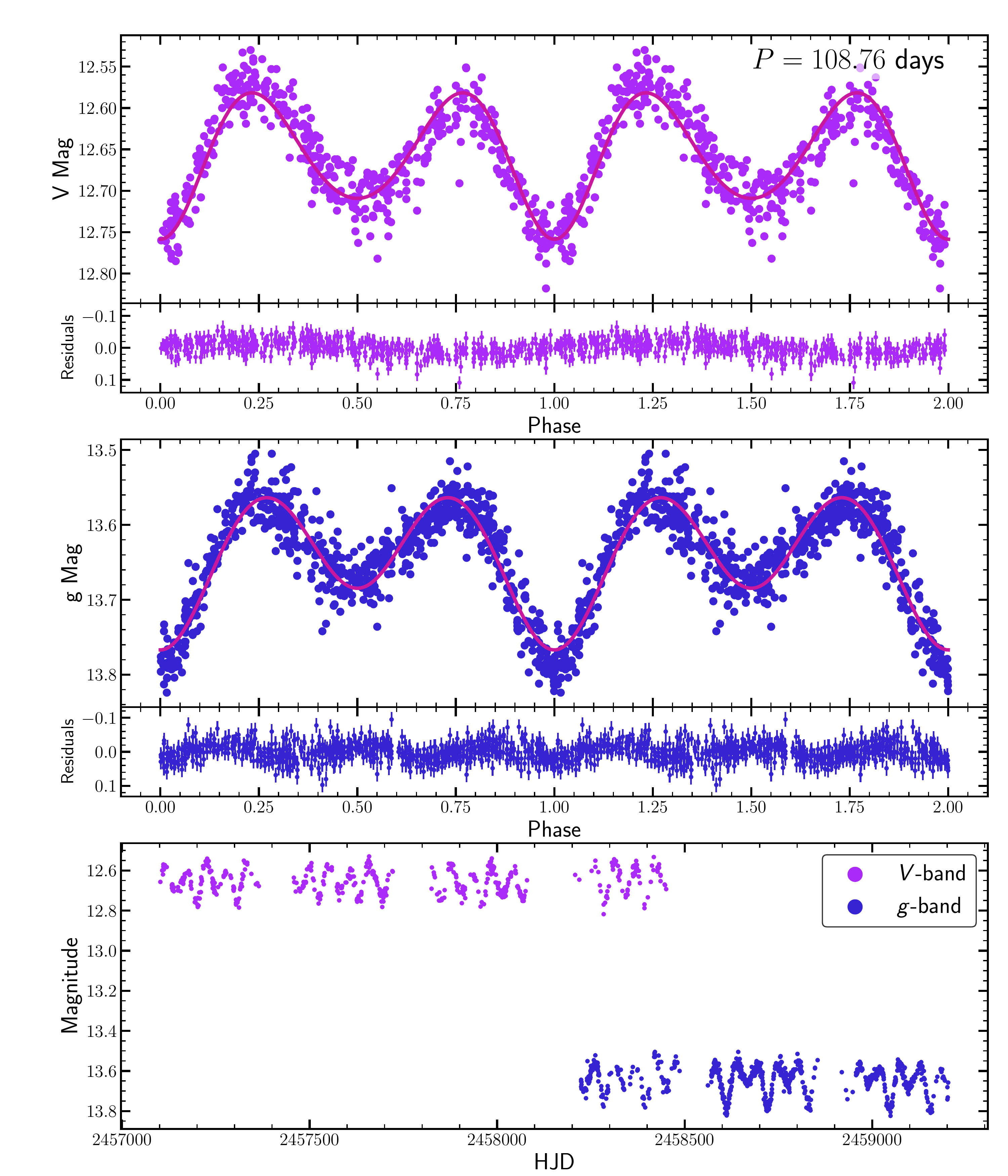}
    \caption{ASAS-SN $V$-band (top), $g$-band (middle), and unfolded (bottom) light curves for ASASSN-V J210204.31$+$394834.5. Solid lines in the light curves correspond to least squares ELL fit. This light curve was selected as part of our visual inspection sample because of the uneven minima, but it is probably an RV Tauri variable.}
    \label{fig:rva_example}
\end{figure}

At longer periods, RV Tauri stars are a potential source of false positives. RV Tauri variables are pulsating post-asymptotic branch or post-red giant branch stars that follow a known period-luminosity relation \citep{Bodi19}. Compared to eclipsing binary light curves, RV Tauri variables are expected to show minima that are both asymmetric and uneven. Figure \ref{fig:rva_example} shows a likely RV Tauri variable (ASASSN-V J210204.31$+$394834.5). In this case, the asymmetric minima and uneven maxima both indicate that ASASSN-V J210204.31$+$394834.5 is unlikely to be an ELL. 

After visually inspecting the stars selected by the red line in Figure \ref{fig:chi_compare}, we repeated the visual inspection process for a second region of the $R$--\chiELL{} parameter space indicated by the dotted line to assess the effectiveness of this selection method. Since many more variables fall in this new parameter space, we restricted this sample to variables previously classified as ROT and SR. We find that the majority of our ELL candidates are found in our original selection region, confirming that the ratio of \chiELL{} to \chiCOS{} is an effective metric to select ELL variables. After multiple rounds of visual inspection we identified a total of \nELL{} ELL candidates and an additional \nECL{} ELLs with eclipses (ELL+ECL) discussed below.

\begin{figure*}
    \centering
    \includegraphics[scale=0.3]{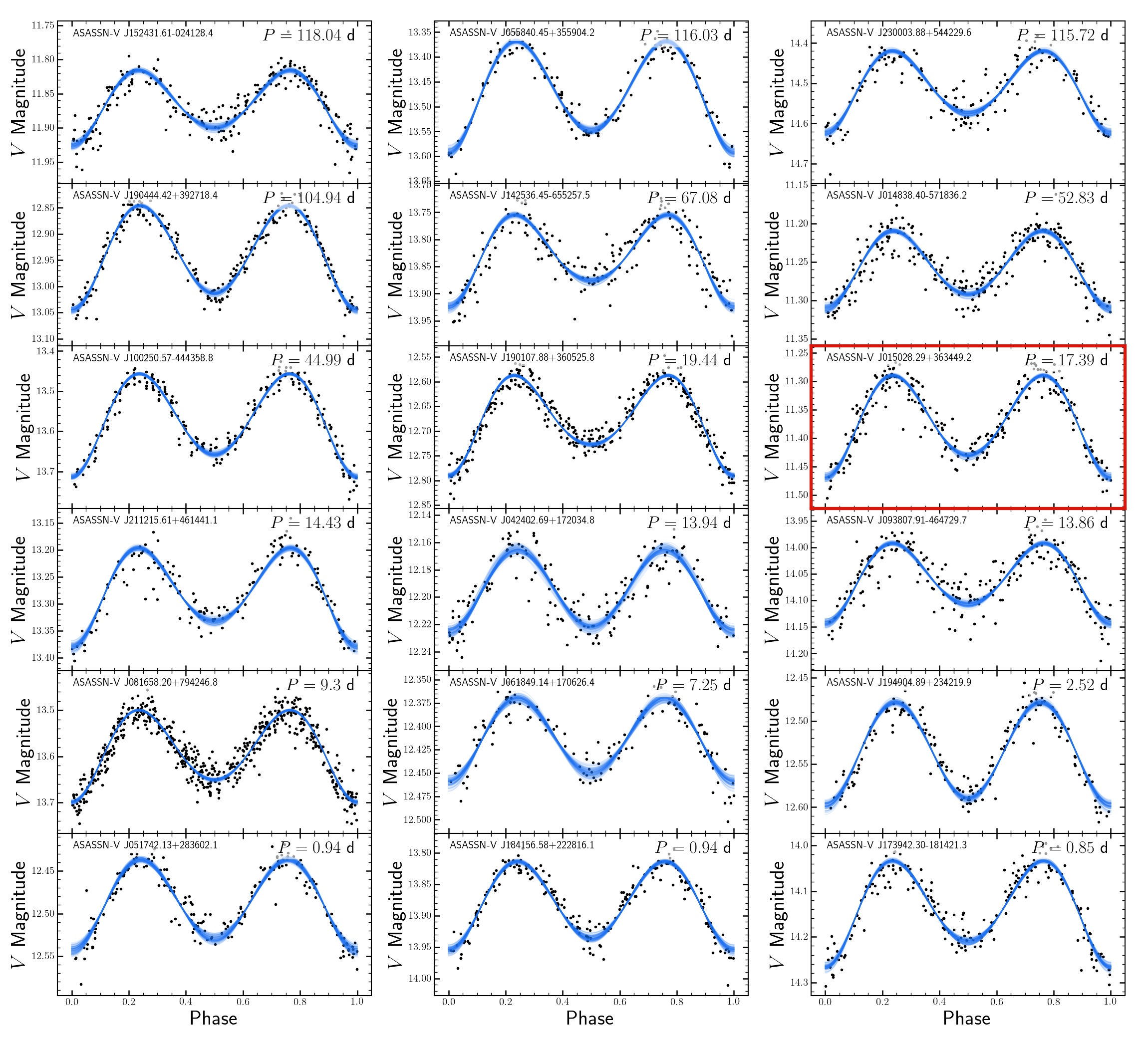}
    \caption{18 randomly selected ELL candidate $V$-band light curves sorted by period. The period is given in upper right corner of each panel. 100 random samples of the MCMC posteriors computed are shown in blue. The MCMC corner plot for the panel highlighted in red (ASASSN-V J015028.29+363449.2) is shown in Figure \ref{fig:example_corner}.}
    \label{fig:RandomLCs}
\end{figure*}

\begin{figure}
    \centering
    \includegraphics[width=\linewidth]{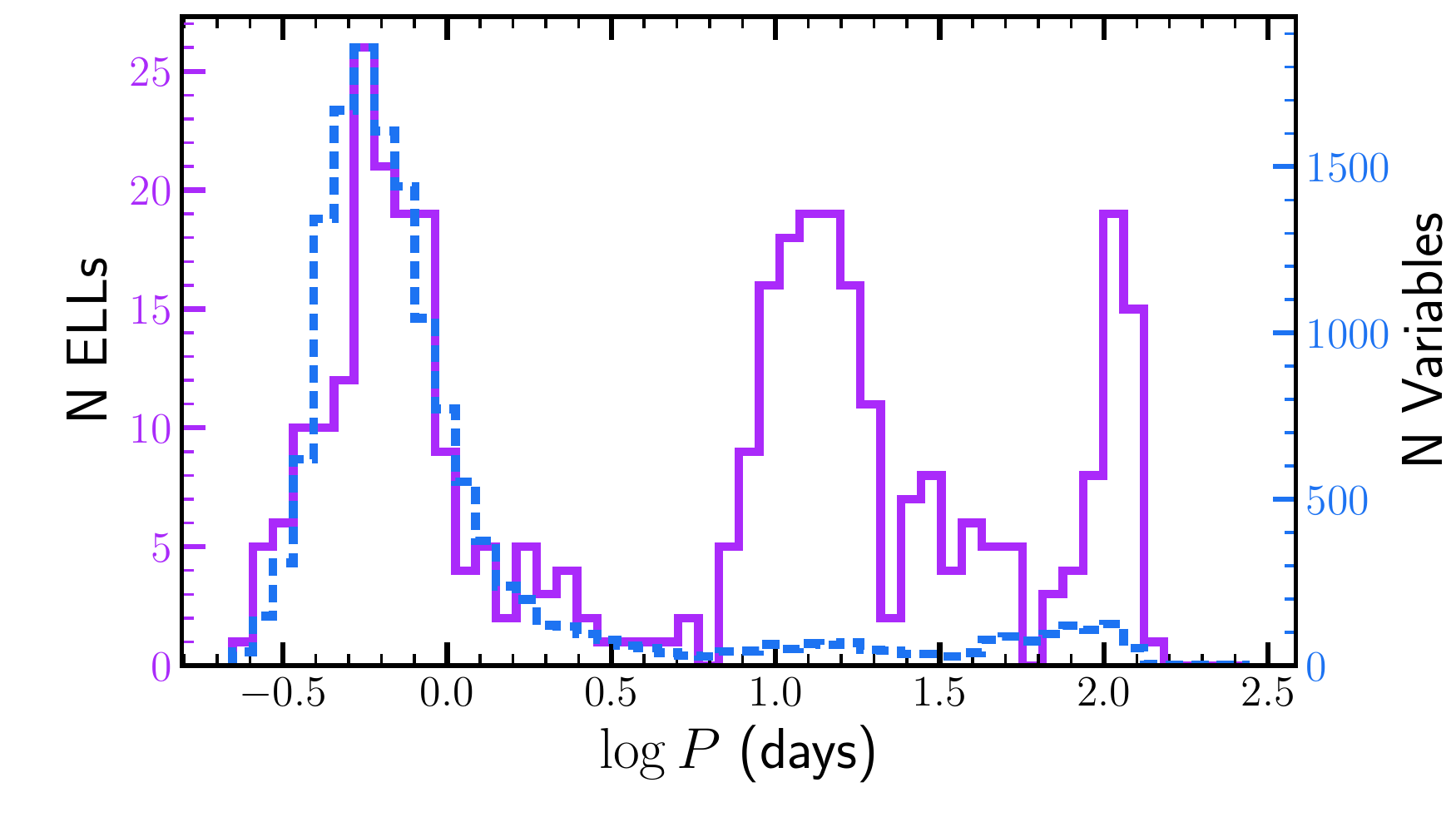}
    \caption{Period distribution for the ELL candidates (purple) and the initial candidates that were selected by the \chiELL{} cut (blue). Periods range form \Pmin{} to \Pmax{} days. The median period is \Pmed{} days. While the period distribution of the initial candidates contains a single peak, we find a triple peaked distribution of ELL periods. These peaks correspond to different evolutionary stages as shown on the CMD in Figure \ref{fig:gaia_cmd}.}
    \label{fig:period_distribution}
\end{figure}

\begin{figure}
    \centering
    \includegraphics[width=\linewidth]{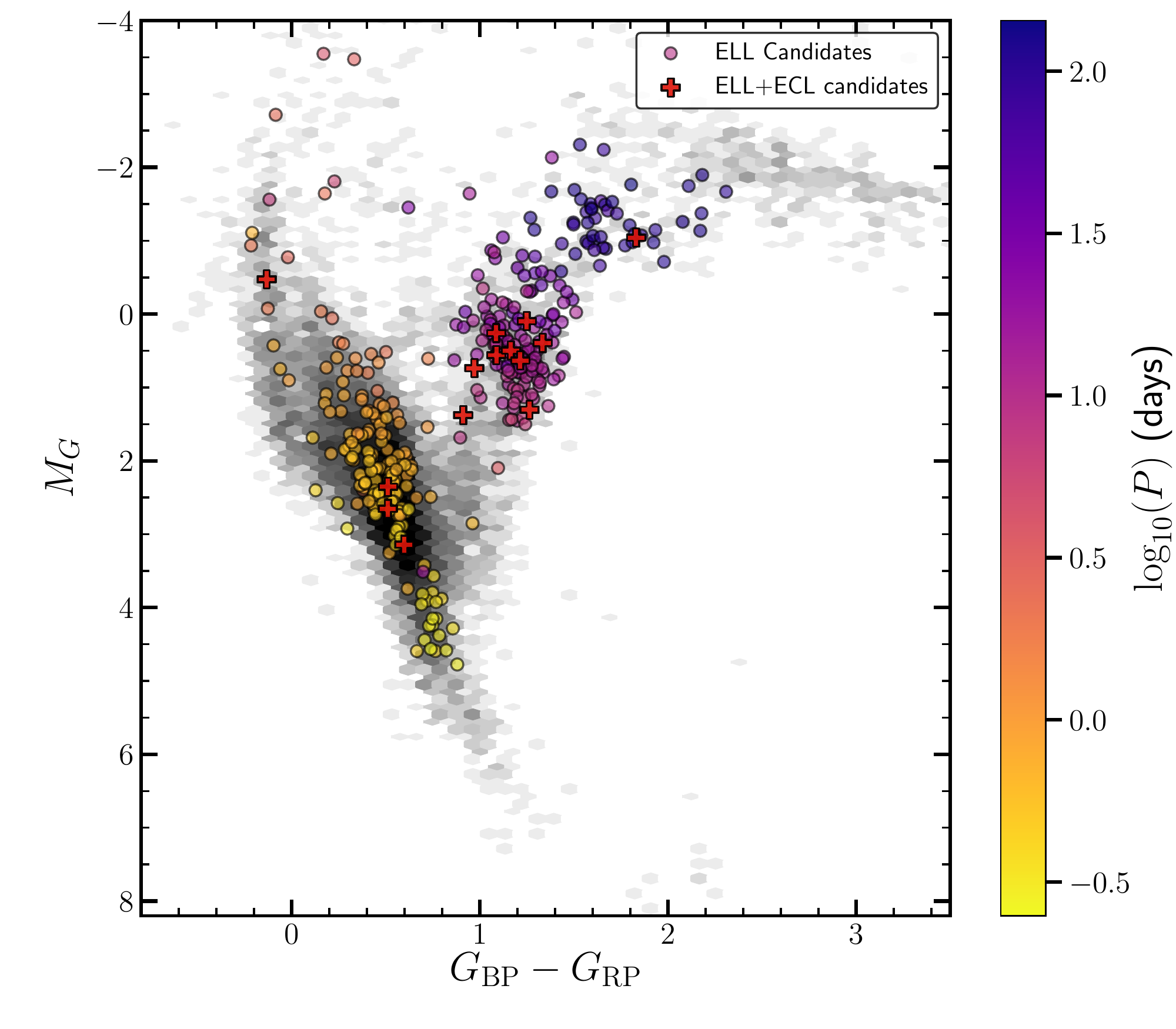}
    \caption{ELL and ELL+ECL variables on a \textit{Gaia} CMD using the \texttt{StarHorse} extinction estimates. All of the variables that were visually inspected are shown in gray and the final ELL candidates are colored by $\log P$. The three peaks in the period distribution of Figure \ref{fig:period_distribution} generally correspond to different evolutionary stages. ELL+ECL variables are shown in red and are not colored by period.}
    \label{fig:gaia_cmd}
\end{figure}

The candidates are listed in Table \ref{tab:ELLTable} and Figure \ref{fig:RandomLCs} shows examples of their $V$-band light curves. Figure \ref{fig:period_distribution} shows the period distribution for the ELL candidates. ELL candidates have periods ranging from \Pmin{} to \Pmax{} days with a median period of \Pmed{} days. While the sample before visual inspection has a single peak at $P\sim0.6$ days, the final ELL candidate distribution shows three peaks. This distribution is consistent with the sample of OGLE ellipsoidal variables in the Galactic bulge \citep{Soszynski16,Gomel21}. The three peaks likely correspond to the conditions for observing ellipsoidal variability at different evolutionary stages. Figure \ref{fig:gaia_cmd} shows the ELL and ELL+ECL candidates on a \textit{Gaia} color-magnitude diagram (CMD) with extinctions from \texttt{StarHorse} \citep{Anders19}. Whereas tidal distortion of main sequence stars can only produce observable ellipsoidal variability for systems with short-period orbits, red clump and giant branch stars can be distorted by companions in much larger orbits. 

\begin{table*}
\caption{18 entries of the electronic table for ELL candidates. The ephemeris is defined for the photometric minimum at phase $\phi=0$. The amplitude is measured between the $5^{\rm{th}}$ and $95^{\rm{th}}$ percentiles. We calculate $e_2$, $M_0$, and the minimum companion mass $M_{c,\rm{min}}$ in Section \S\ref{sec:minimum_companion_mass}. The light curves are shown in Figure \ref{fig:RandomLCs}.}
\label{tab:ELLTable}
\begin{center}
\begin{threeparttable}
\begin{tabular}{lS[table-format=2.2]rS[table-format=2.2]rrS[table-format=2.2]S[table-format=2.2]S[table-format=2.2]S[table-format=2.2]}
\toprule
      {ASAS-SN Name} &  {Period (days)} &  {Ephemeris} &  {V Mag} &  {Amplitude (mag)} &  {$M_*\ (M_\odot)$} & {$R_*\ (R_\odot)$} &  {$e_2$} &  {$M_0\ (M_\odot)$} &  {$M_{c,\rm{min}}\ (M_\odot)$} \\
\midrule
 J152431.61-024128.4 &           118.04 &    2456591.8 &    11.87 &               0.12 &            1.02 &              51.82 &    0.026 &                5.03 &                            0.21 \\
 J055840.45+355904.2 &           116.03 &    2456925.8 &    13.47 &               0.23 &            1.17 &              57.31 &    0.053 &                3.41 &                            0.40 \\
 J230003.88+544229.6 &           115.72 &    2456956.6 &    14.51 &               0.22 &            1.09 &              47.31 &    0.047 &                2.21 &                            0.53 \\
 J190444.42+392718.4 &           104.94 &    2456515.9 &    12.93 &               0.20 &            1.14 &              43.98 &    0.047 &                2.12 &                            0.61 \\
 J142536.45-655257.5 &            67.08 &    2457429.9 &    13.83 &               0.19 &            1.28 &              61.50 &    0.037 &               18.03 &                            0.09 \\
 J014838.40-571836.2 &            52.83 &    2456779.1 &    11.25 &               0.11 &            1.08 &              19.93 &    0.024 &                1.52 &                            0.76 \\
 J100250.57-444358.8 &            44.99 &    2457388.1 &    13.56 &               0.25 &            1.03 &              21.57 &    0.056 &                1.14 &                            0.93 \\
 J190107.88+360525.8 &            19.44 &    2456582.2 &    12.67 &               0.21 &            1.22 &              17.53 &    0.041 &                4.52 &                            0.33 \\
 J015028.29+363449.2 &            17.39 &    2456634.9 &    11.37 &               0.18 &            1.09 &              14.16 &    0.042 &                2.94 &                            0.40 \\
 J211215.61+461441.1 &            14.43 &    2457098.6 &    13.28 &               0.20 &            1.19 &              20.07 &    0.039 &               12.84 &                            0.11 \\
 J042402.69+172034.8 &            13.94 &    2456224.9 &    12.20 &               0.07 &            1.07 &              12.59 &    0.020 &                6.79 &                            0.17 \\
 J093807.91-464729.7 &            13.86 &    2457414.4 &    14.06 &               0.17 &            1.08 &              15.93 &    0.034 &                7.98 &                            0.15 \\
 J081658.20+794246.8 &             9.30 &    2456678.9 &    13.59 &               0.20 &            1.06 &               8.39 &    0.044 &                2.00 &                            0.56 \\
 J061849.14+170626.4 &             7.25 &    2457003.0 &    12.41 &               0.10 &            3.83 &              15.28 &    0.031 &               28.55 &                            0.51 \\
 J194904.89+234219.9 &             2.52 &    2457077.7 &    12.53 &               0.13 &            5.60 &    26.50\tnote{$\dagger$} &    0.043 &              890.05 &                            0.04 \\
 J051742.13+283602.1 &             0.94 &    2457007.6 &    12.49 &               0.11 &            2.09 &               4.43 &    0.033 &               38.40 &                            0.11 \\
 J184156.58+222816.1 &             0.94 &    2457914.6 &    13.88 &               0.15 &            1.75 &               4.02 &    0.049 &               19.63 &                            0.16 \\
 J173942.30-181421.3 &             0.85 &    2457070.6 &    14.13 &               0.22 &            1.60 &               3.56 &    0.071 &               11.46 &                            0.22 \\
\bottomrule
\end{tabular}

\begin{tablenotes}
\item[$\dagger$] Radius derived from the \texttt{StarHorse} effective temperature, surface gravity, and metallicity is unphysical given the orbital period and estimated mass.
\end{tablenotes}
\end{threeparttable}
\end{center}
\end{table*}

\subsection{Ellipsoidal variables with eclipses} \label{sec:ELL+ECL}

\begin{figure*}
    \centering
    \includegraphics[width=\linewidth]{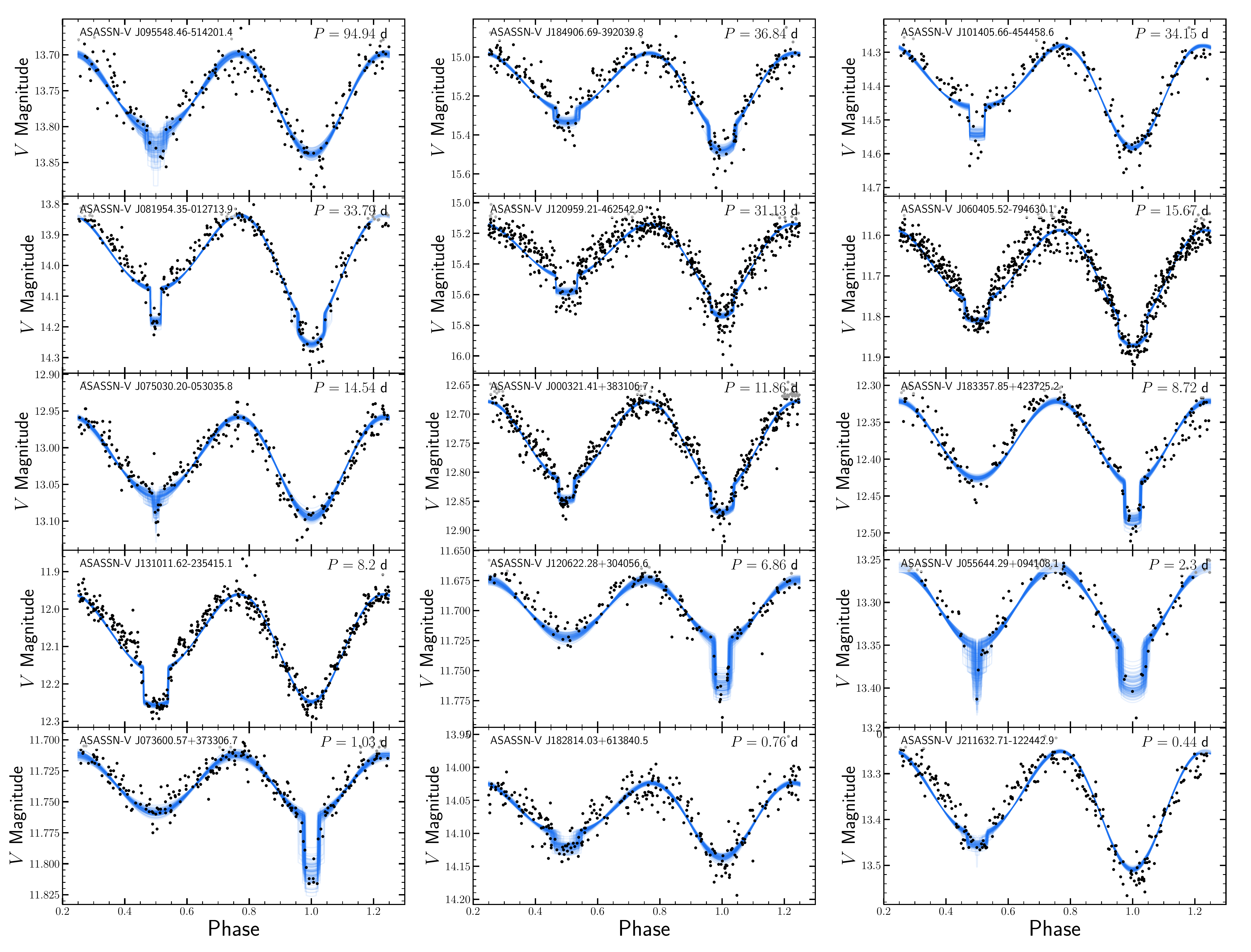}
    \caption{Light curves of \nECL{} candidates that show eclipses in addition to the ellipsoidal modulations. 100 random samples of the MCMC posteriors are shown in blue. We use the same definition of phase, where $\phi=0$ corresponds to the ELL minimum, but adjust the range of the axis to better illustrate the eclipses. Eclipse depth and duration are given in Table \ref{tab:ELLECL}.} 
    \label{fig:ellecl}
\end{figure*}

During visual inspection we identified \nECL{} variables that show both eclipses and ellipsoidal modulations (ELL+ECL). The phase of the binary is well constrained when fitting the analytical model in Equation \ref{eqn:analytical model}, so we expect any eclipses to occur at phase $\phi=0$, when the companion eclipses the photometric primary, or at phase $\phi=0.5$, when the photometric primary eclipses the companion. For the systems that we flagged as having eclipsing features at $\phi=0$ and/or $\phi=0.5$, we modified the ELL analytic model to include one or two boxcar components. The ELL+ECL model is fit using Markov Chain Monte Carlo (MCMC) methods with the Python package \texttt{emcee} \citep{Foreman-Mackey13}. The resulting eclipse width is used to estimate the eclipse duration given the orbital period. The eclipse duration and depth are included in Table \ref{tab:ELLECL} and Figure \ref{fig:ellecl} shows the \nECL{} light curves with the modified ELL+ECL fit. 

As compared to the ELL candidates, ELL+ECL systems typically have longer periods, with a median period of \eclPmed{} days, and slightly higher \chiELL{} (Figure \ref{fig:chi_compare}). The majority of ELL+ECL systems are also on the giant branch (Figure \ref{fig:gaia_cmd}). We are more successful at finding ELL+ECL systems at longer periods as compared to the ELL candidates. 

\begin{table*}
\centering
\caption{Ellipsoidal variables with eclipses (ELL+ECL) sorted by period. We estimate the eclipse depth and duration by adding one or two boxcar functions to the ELL analytical model. For sources where there are two eclipsing features, we designate the primary eclipse to correspond to the deeper ELL model minimum.}
\label{tab:ELLECL}
\resizebox{\textwidth}{!}{\begin{tabular}{lS[table-format=2.2]rS[table-format=2.2]S[table-format=2.2]rrrr}
\toprule
      {ASAS-SN Name} &  {Period (days)} &  {Ephemeris} &  {V Mag} &  {Amplitude (mag)} &      {Eclipse Depth (mag)} &  {Eclipse Duration (days)} &    {Secondary Depth (mag)} & {Secondary Duration (days)} \\
\midrule
 J095548.46-514201.4 &            94.94 &    2457349.4 &    13.76 &               0.15 &  $0.021^{+0.033}_{-0.009}$ &  $5.498^{+8.375}_{-3.684}$ &                            &                             \\
 J184906.69-392039.8 &            36.84 &    2456775.8 &    15.15 &               0.50 &  $0.095^{+0.111}_{-0.079}$ &  $3.128^{+3.416}_{-2.850}$ &  $0.056^{+0.071}_{-0.041}$ &   $2.581^{+3.126}_{-2.195}$ \\
 J101405.66-454458.6 &            34.15 &    2457422.3 &    14.41 &               0.31 &  $0.081^{+0.092}_{-0.070}$ &  $1.678^{+1.699}_{-1.656}$ &                            &                             \\
 J081954.35-012713.9 &            33.79 &    2456594.7 &    13.98 &               0.40 &  $0.066^{+0.073}_{-0.058}$ &  $2.903^{+3.158}_{-2.682}$ &  $0.105^{+0.115}_{-0.096}$ &   $1.157^{+1.200}_{-1.114}$ \\
 J120959.21-462542.9 &            31.13 &    2456765.0 &    15.35 &               0.62 &  $0.094^{+0.106}_{-0.082}$ &  $2.352^{+2.382}_{-2.338}$ &  $0.101^{+0.114}_{-0.087}$ &   $2.120^{+2.166}_{-2.096}$ \\
 J060405.52-794630.1 &            15.67 &    2456778.3 &    11.69 &               0.29 &  $0.047^{+0.051}_{-0.043}$ &  $1.450^{+1.460}_{-1.440}$ &  $0.043^{+0.047}_{-0.039}$ &   $1.181^{+1.194}_{-1.170}$ \\
 J075030.20-053035.8 &            14.54 &    2456597.4 &    13.02 &               0.14 &  $0.018^{+0.034}_{-0.006}$ &  $0.316^{+0.762}_{-0.097}$ &                            &                             \\
 J000321.41+383106.7 &            11.86 &    2456596.8 &    12.75 &               0.19 &  $0.038^{+0.042}_{-0.033}$ &  $0.862^{+0.938}_{-0.818}$ &  $0.033^{+0.038}_{-0.028}$ &   $0.576^{+0.645}_{-0.531}$ \\
 J183357.85+423725.2 &             8.72 &    2456592.2 &    12.37 &               0.16 &  $0.049^{+0.055}_{-0.043}$ &  $0.445^{+0.494}_{-0.395}$ &                            &                             \\
 J131011.62-235415.1 &             8.20 &    2456788.4 &    12.08 &               0.31 &  $0.091^{+0.096}_{-0.086}$ &  $0.649^{+0.669}_{-0.629}$ &                            &                             \\
 J120622.28+304056.6 &             6.86 &    2456780.5 &    11.70 &               0.08 &  $0.036^{+0.043}_{-0.030}$ &  $0.344^{+0.421}_{-0.289}$ &                            &                             \\
 J055644.29+094108.1 &             2.30 &    2457036.9 &    13.30 &               0.13 &  $0.044^{+0.054}_{-0.034}$ &  $0.186^{+0.210}_{-0.151}$ &  $0.041^{+0.062}_{-0.022}$ &   $0.052^{+0.133}_{-0.020}$ \\
 J073600.57+373306.7 &             1.03 &    2456615.9 &    11.74 &               0.07 &  $0.043^{+0.050}_{-0.035}$ &  $0.050^{+0.060}_{-0.043}$ &                            &                             \\
 J182814.03+613840.5 &             0.76 &    2456674.5 &    14.07 &               0.13 &  $0.020^{+0.027}_{-0.013}$ &  $0.062^{+0.073}_{-0.043}$ &                            &                             \\
 J211632.71-122442.9 &             0.44 &    2456236.6 &    13.35 &               0.20 &  $0.022^{+0.030}_{-0.014}$ &  $0.028^{+0.031}_{-0.024}$ &                            &                             \\
\bottomrule
\end{tabular}
}
\end{table*}

\subsection{Minimum Companion Mass} \label{sec:minimum_companion_mass}

Without radial velocity data we cannot fully confirm the ELL nature of the candidates or the properties of the companion stars. However, reliable parallaxes from \textit{Gaia} Early Data Release 3 \citep[EDR3, ][]{GaiaCollaboration16, GaiaCollab20}  and photometry have been used in both \textit{Gaia} Data Release 2 \citep[DR2, ][]{GaiaCollaboration18} and \texttt{StarHorse} \citep{Querioz18, Anders19} to estimate the stellar parameters.

The Gaia DR2 catalog has luminosity, temperature, and radius measurements for \nGaiaR{} of the ELL candidates. These radii are computed using APSIS-Flame assuming an extinction of $A_G=0$ \citep{Andrae18}. Better estimates can be obtained from \texttt{StarHorse} \citep{Querioz18} which uses a broader range of photometry to fit the stellar spectral energy distributions (SEDs) and then estimate masses, temperatures, surface gravities, and extinctions. Figure \ref{fig:gaia_cmd} shows a CMD of the \nSH{} ELL candidates in the \citet{Anders19} catalog. We used \texttt{Isoclassify} \citep{Huber17,Berger20} to interpolate over the \texttt{MIST} \citep{Dotter16, Choi16, Paxton11} model bolometric corrections to calculate the luminosity of each star given the estimated extinction, temperature, distance, and $G$-band magnitude. This allows us to calculate the stellar radius $R_*$ given the \citet{Anders19} $T_{\rm{eff}}$ measurements.

Next, we refit the $V$-band light curves of the candidates in terms of the underlying physical parameters in Equation \ref{eqn:analytical model} using MCMC with \texttt{emcee} \citep{Foreman-Mackey13, Foreman-Mackey2016}. We use the $V$-band light curves rather than the $g$-band because of the longer baseline of observation. For each variable we perform 50,000 iterations with 200 walkers. Initial positions are selected based on a least squares fit. We sample over inclination, mass ratio, and $(R_*/a)$. The photospheric parameters $u$ and $\tau$ are held fixed from a linear interpolation over the \citet{Claret11} tables given the  effective temperature, surface gravity, and metallicity estimates. We use a value of 2.0 km/s for the microturbulent velocity. 

\begin{figure}
    \centering
    \includegraphics[width=\linewidth]{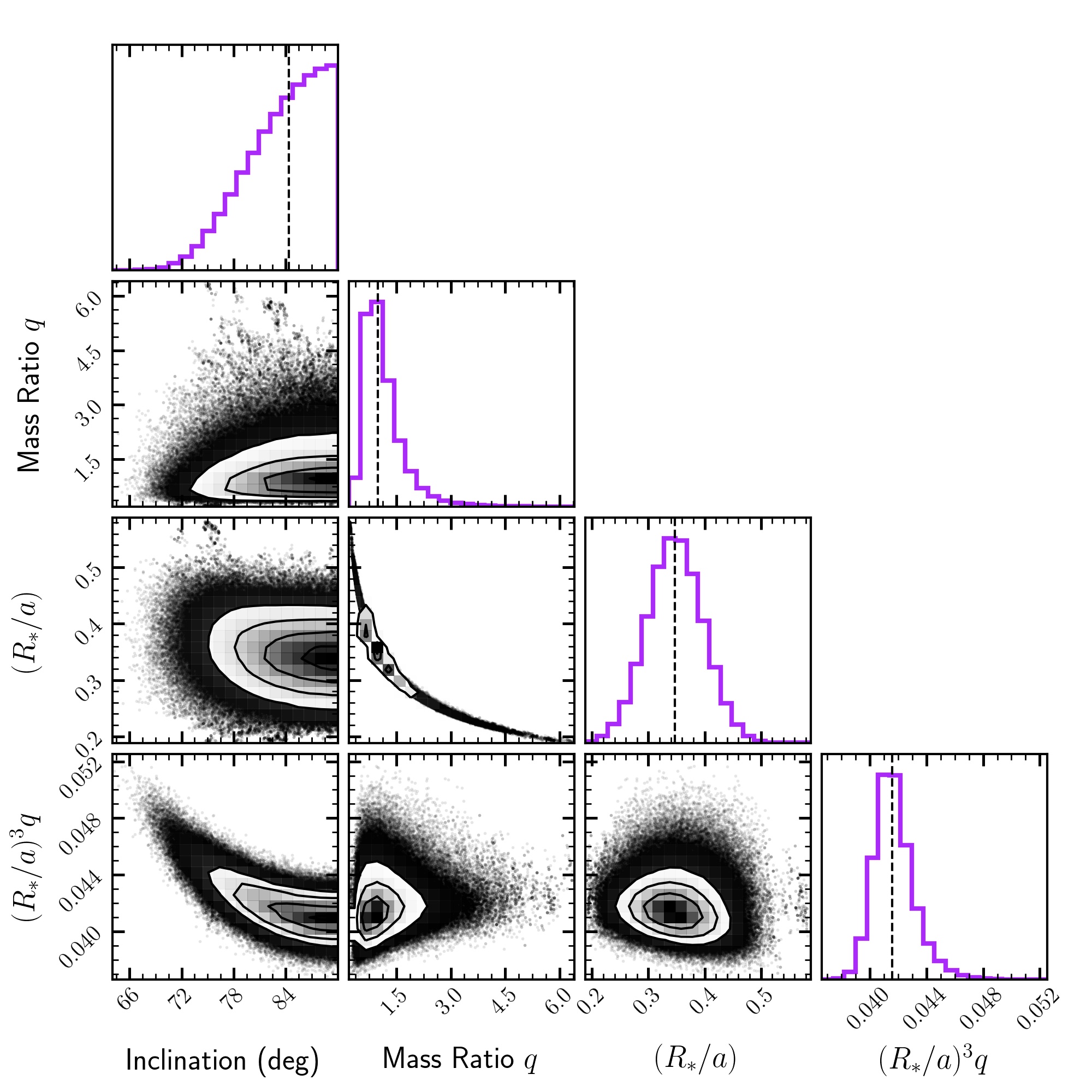}
    \caption{MCMC posteriors for ASASSN-V J015028.29+363449.2. The vertical lines on the histograms show the median value. We use the posterior on $e_2=(R_*/a)^3q$ to sort ELL candidates. The light curve of this system is shown in the highlighted panel of Figure \ref{fig:RandomLCs}.}
    \label{fig:example_corner}
\end{figure}

Figure \ref{fig:example_corner} shows the results for candidate ASASSN-V J015028.29+363449.2, whose light curve is shown in Figure \ref{fig:RandomLCs}. The best-measured quantity related to the masses is the amplitude of the~$\cos 2\theta$ Fourier term
\begin{equation}
    e_2 = q(R_*/a)^3.
\end{equation}
The separate values of $q$ and $R_*/a$, while constrained by the amplitudes of the other Fourier terms, are relatively degenerate. The \texttt{emcee} sampler is affine-invariant and appears to have little difficulty sampling the $q$-$R_*/a$ parameter space despite the near degeneracy. We experimented with sampling in $e_2$ and either $q$ or $R_*/a$ but it appeared to make no practical difference.

Given the period $P$ of the binary and the photometric estimate of the radius of the primary $R_*$, the total binary mass is
\begin{equation}
    M_T = M_c+M_*= \frac{4\pi^2 R_*^3 q}{G P^2 e_2}
\end{equation}
where $M_c$ is the companion mass and $M_*$ is the primary mass. Since the estimates of $q$ are fairly degenerate with estimates of $R_*/a$, the quantity
\begin{equation}
    M_0 = \frac{M_T}{q} = \frac{4\pi^2 R_*^3}{G P^2 e_2}
\end{equation}
is the mass scale most reliably measured by the light curve models. In terms of $M_*$ and $M_0$, the companion mass is
\begin{equation}\label{eqn:mc0}
    M_c = \frac{M_*^2}{M_0-M_*}
\end{equation}
The difference in the denominator tends to be very uncertain, so we focus on a more robust lower limit on the companion mass
\begin{equation}
    M_c > M_{c,\rm{min}} = \frac{M_*^2}{M_0},
\end{equation}

\begin{figure}
    \centering
    \includegraphics[width=\linewidth]{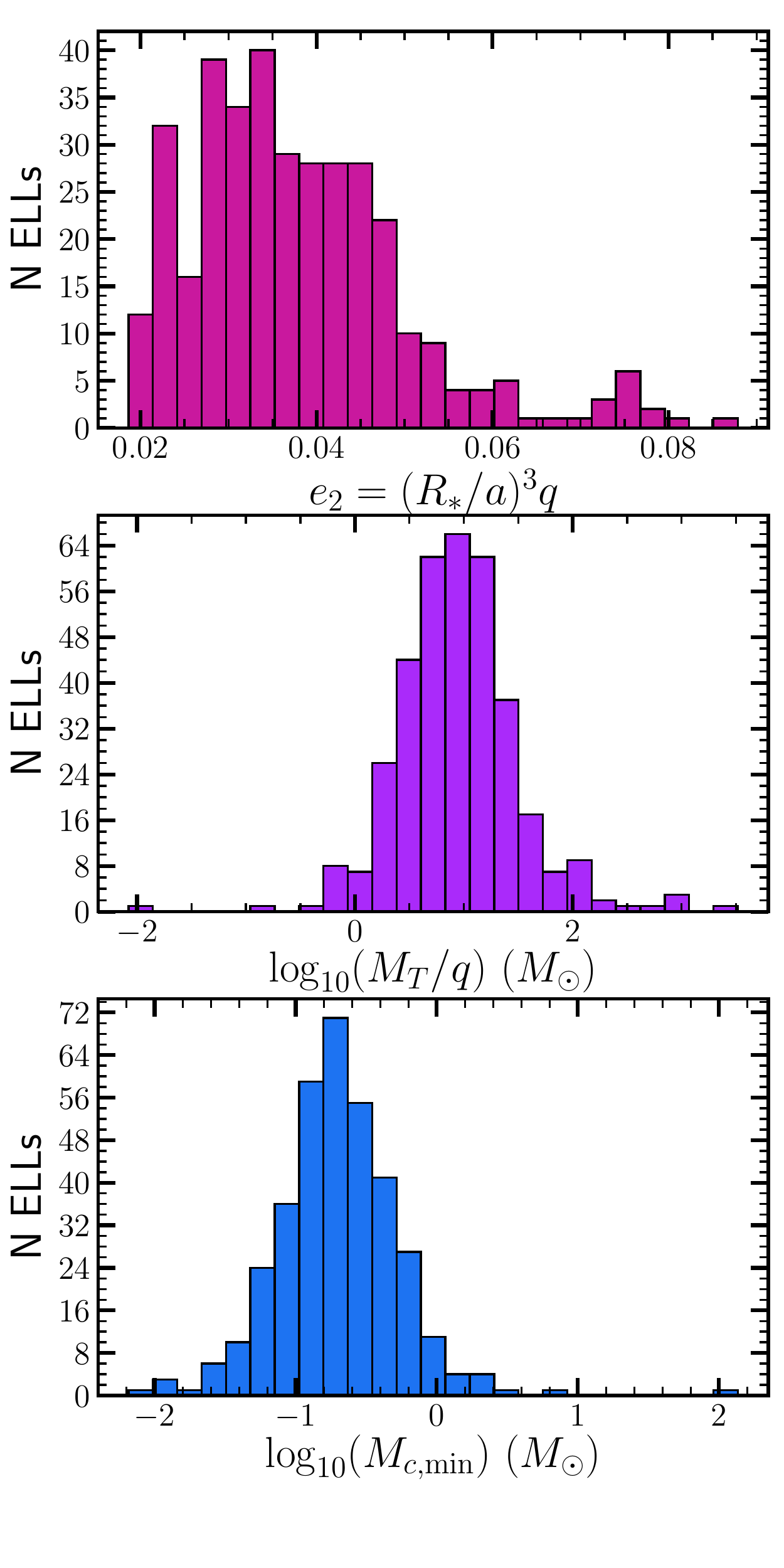}
    \caption{Distributions of $e_2$, $M_T/q$ and $M_{c,\rm{min}}$ for ELL candidates with {\tt StarHorse} estimates of the stellar properties.}
    \label{fig:SH_hists}
\end{figure}

Figure \ref{fig:SH_hists} shows the distributions of the median posteriors for $e_2$, $M_0$, and $M_{c, \rm{min}}$. We find the median minimum companion mass is \McminMed{} $M_\odot$ and \nMCgtONE{} systems have minimum companion masses greater than 1 $M_\odot$. In the absence of RV measurements, the minimum companion mass can be used to prioritize follow-up observations to search for non-interacting compact object binaries. ASAS-SN J095846.87-443947.1 has $M_{c,\ \rm{min}}=136\ M_\odot$, the largest in the candidate catalog. While this outlying minimum mass estimate probably indicates that the main sequence star is not an ELL, we do not reject any variables based on the mass estimates calculated from the photometric fits. 

We can also use known non-interacting compact object binary systems to place an additional check on our sorting method. We fit the V723~Mon KELT $R$-band light curve shown in Figure \ref{fig:unicorn} with the ELL analytical model. Using the fit $e_2=0.0493$, the {\tt StarHorse} $M_*=1.09\ M_{\odot}$, and $R=22.70\ R_{\odot}$, we find the minimum companion mass is $M_c > 1.39\ M_{\odot}$. Not only is this consistent with the mass measured by \citet{Jayasinghe21}, but it is also higher than almost all ELL candidates in our catalog, suggesting that such systems are rare. 

We find \SHinvalidR{} ELL candidates where the \texttt{StarHorse} radius estimates are unphysical given the orbital period of the binary. ASAS-SN J193909.43+232051.4, for example, has an orbital period of $P=4.37$ days and a radius estimate of $R=53.71\ R_{\odot}$. This is the most luminous star on the upper-main sequence shown in Figure \ref{fig:gaia_cmd} at $M_G=-3.55$~mag. All of these stars are within 10 degrees of the galactic plane, suggesting increased extinction may result in less reliable {\tt StarHorse} measurements. Stars with such unphysical radii are marked in Table \ref{tab:ELLTable}. Many of the outlying stars in the $M_0$ and $M_{c,\ \rm{min}}$ distributions have inaccurate radii. ASAS-SN J193909.43+232051.4, for example, has $M_0>1000\ M_\odot$ and $M_{c,\ \rm{min}} = 0.006\ M_\odot$.

\section{Radial Velocities and X-rays} \label{sec:rv_xray}

We next search for systems with existing multi-epoch RV measurements. We cross-match our ELL candidate catalog with the Large Sky Area Multi-Object Fibre Spectroscopic Telescope Survey Data Release 5 \citep[LAMOST DR5,][]{Luo15, Yuan15}, the Apache Point Observatory Galactic Evolution Experiement Data Release 16 \citep[APOGEE DR16,][]{Holtzman15, Garcia16, Ahumada20}, and the the Sixth Data Release of the Radial Velocity Experiment \citep[RAVE DR6, ][]{Steinmentz20I, Steinmetz20II} with a matching radius of $5\farcs0$. We find \nAPOGEE{} APOGEE DR16 cross-matches, \nAPOGEEmultiple{} of which have multiple observations. We find \nLAMOST{} LAMOST DR5 cross-matches, \nLAMOSTmultiple{} of which have multiple observations. We find \nRAVE{} RAVE DR6 cross-matches, none of which have multiple observations.

\begin{figure}
    \centering
    \includegraphics[width=\linewidth]{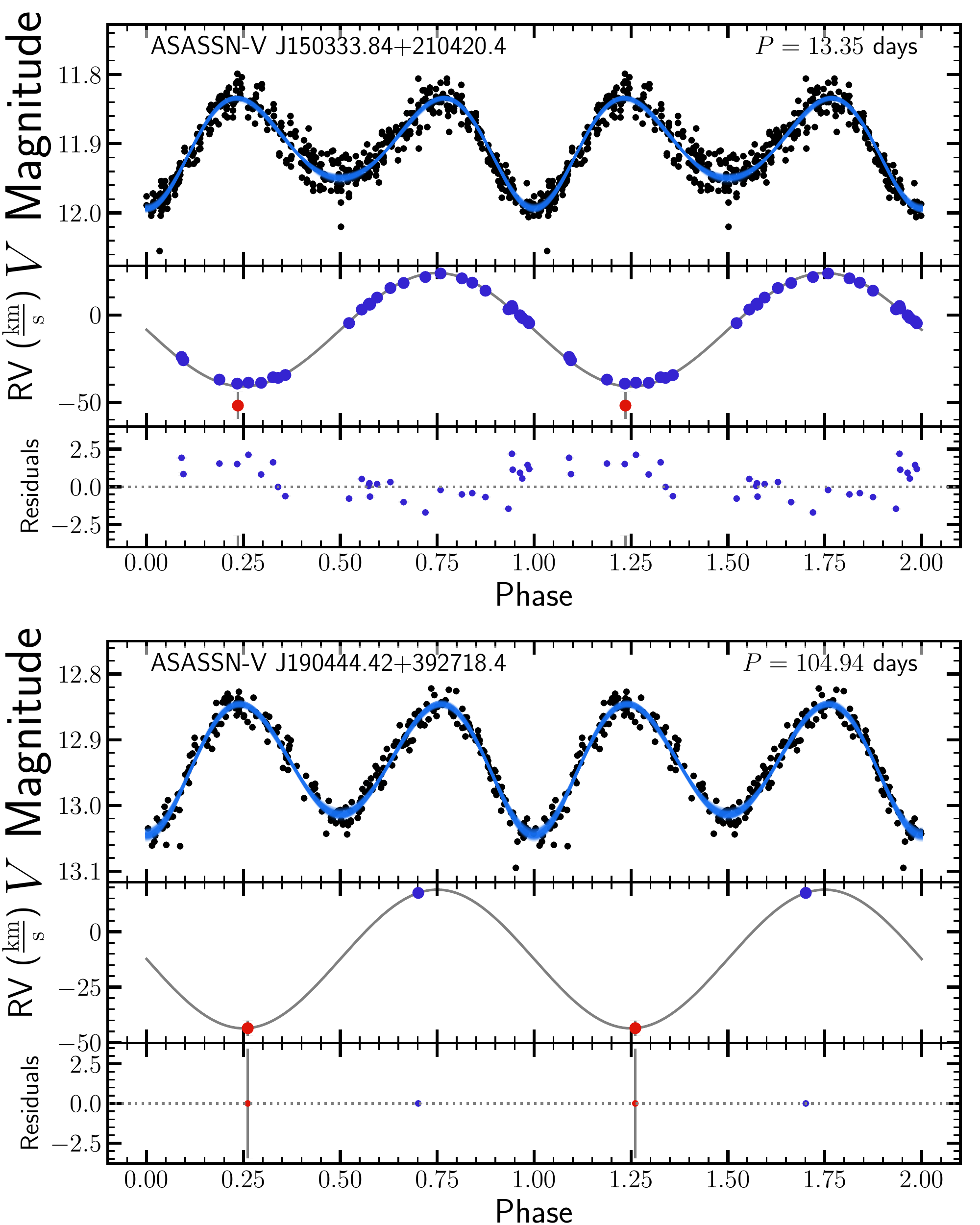}
    \caption{ASAS-SN $V$-band light curves and radial velocity measurements for ASASSN-V J150333.84+210420.4 (top) and J190444.42+392718.4 (bottom). APOGEE DR16 measurements are shown in blue and LAMOST DR5 are in red. The magenta lines show the least squares fit to the ELL analytical model. The gray lines show the best fit circular orbit given the photometric period and phase where $\phi=0$ is defined as the photometric minimum.}
    \label{fig:rv_panel}
\end{figure}

For the systems with multiple RV measurements that have a phase separation $\Delta\phi>0.25$ we fit a circular orbit to the RV data using the photometric period and phase. Since $\phi=0$ corresponds to photometric minimum, the maximum of the radial velocity curve will occur at $\phi=0.75$. We use the resulting semi-amplitude $K$ to calculate the mass function
\begin{equation}
    f(M) = \frac{P K^3}{2\pi G}=\frac{M_c^3 \sin^3 i}{(M_*+M_c)^2}.
\end{equation}
If we again assume the \texttt{StarHorse} mass estimates for $M_*$ we can calculate the mass of the companion $M_c$. The inclination and $\sin^3 i$ posteriors from the MCMC fit to the ELL analytical model are broad for many candidates, such as the example shown in Figure \ref{fig:example_corner}, so we simply use an edge-on inclination to calculate the minimum companion mass for the RV fits. Table \ref{tab:rv_table} lists the ELL candidates with multiple RV observations and the derived amplitudes $K$, mass functions$f$, and $M_c$. Figure \ref{fig:rv_panel} shows the light curves and phased radial velocity observations for two of these systems, ASASSN-V J150333.84+210420.4 and J190444.42+392718.4. None of these systems yield companion masses that are promising candidates for neutron star or black hole companions, although they could be non-interacting white dwarf companions. 

\begin{table*}
\caption{ELL Candidates with Multiple RV observations from APOGEE DR16 and LAMOST DR5 sorted by period. Four have $\rm{Max}(\Delta\phi)>0.25$ but are poorly fit by a circular model so we do not include $K$, $f$, or $M_{c,\rm{RV}}$. The V723~Mon RV results from \citet{Jayasinghe21} are included for comparison.} 
\label{tab:rv_table}
\begin{tabular}{lrrrrrrrr}
\toprule
   {ASAS-SN Name} & {Period (days)} &  {N$_{\rm{RV}}$} & {Max($\Delta\phi$)} & {$K$ (km/s)} & {$f(M)\ (M_{\odot})$} & {$M_{c,\rm{RV}}\ (M_{\odot})$} & {$M_{c,\rm{min}}\ (M_{\odot})$} &           {Survey} \\
\midrule
 J055450.78+242657.3 &             143.19 &                   3 &                   0.41 &            8.75 &                     0.01 &                              0.26 &                               0.07 &                APOGEE \\
 J041857.27+400108.6 &             119.43 &                   2 &                   0.03 &                 &                          &                                   &                                    &                APOGEE \\
 J152431.61-024128.4 &             118.04 &                   3 &                   0.31 &           12.53 &                     0.02 &                              0.36 &                               0.21 &          RAVE, LAMOST \\
 J114807.73-020920.5 &             113.42 &                   2 &                   0.46 &           22.64 &                     0.14 &                              0.70 &                               0.17 &                LAMOST \\
 J190444.42+392718.4 &             104.94 &                   2 &                   0.44 &           31.29 &                     0.33 &                              1.23 &                               0.61 &        APOGEE, LAMOST \\
 J221959.83+673236.6 &              76.69 &                   2 &                   0.04 &                 &                          &                                   &                                    &                APOGEE \\
 J093300.84+341709.2 &              44.68 &                   5 &                   0.16 &                 &                          &                                   &                                    &        APOGEE, LAMOST \\
 J140925.52+512653.7 &              29.97 &                   3 &                   0.06 &                 &                          &                                   &                                    &                LAMOST \\
 J001532.90+384119.9 &              27.56 &                   4 &                   0.34 &           48.28 &                     0.32 &                              1.17 &                               0.81 &                LAMOST \\
 J032653.26+005506.7 &              19.32 &                   5 &                   0.26 &           33.17 &                     0.07 &                              0.60 &                               0.15 &                APOGEE \\
 J042402.69+172034.8 &              13.94 &                   4 &                   0.36 &           64.44 &                     0.39 &                              1.29 &                               0.17 &        APOGEE, LAMOST \\
 J150333.84+210420.4 &              13.35 &                  30 &                   0.50 &           32.53 &                     0.05 &                              0.49 &                               0.39 &        APOGEE, LAMOST \\
 J081658.20+794246.8 &               9.30 &                   3 &                   0.38 &           47.86 &                     0.11 &                              0.69 &                               0.56 &                APOGEE \\
 J065624.05+245825.6 &               4.81 &                   3 &                   0.23 &                 &                          &                                   &                                    &                APOGEE \\
 J062252.43+034920.5 &               0.77 &                   4 &                   0.00 &                 &                          &                                   &                                    &                LAMOST \\
 J075654.91+474622.9 &               0.66 &                   3 &                   0.40 &           90.05 &                     0.05 &                              0.58 &                               0.57 &                LAMOST \\
 J101909.73+414611.1 &               0.45 &                   3 &                   0.45 &            8.96 &                 $<0.01$ &                              0.04 &                               0.14 &                LAMOST \\
 J150355.74-011623.2 &               0.43 &                   2 &                   0.00 &                 &                          &                                   &                                    &                LAMOST \\
 J081929.94+090258.3 &               0.37 &                   2 &                   0.21 &                 &                          &                                   &                                    &                LAMOST \\
\hline J080826.59-055109.9 &               1.19 &                   2 &                   0.38 &                 &                          &                                   &                                    &                LAMOST \\
 J084328.53+402247.5 &               1.02 &                   4 &                   0.40 &                 &                          &                                   &                                    &                LAMOST \\
 J061635.56+231909.4 &               0.87 &                   3 &                   0.31 &                 &                          &                                   &                                    &                LAMOST \\
 J161803.43+420416.8 &               0.32 &                   4 &                   0.48 &                 &                          &                                   &                                    &                APOGEE \\
\hline            V723 Mon &              60.04 &                 128 &                    0.5 &           65.15 &                     1.72 &                              3.04 &                               1.39 &  \citep{Jayasinghe21} \\
\bottomrule
\end{tabular}

\end{table*}

ASASSN-V J150333.84+210420.4 has the most radial velocity points (29 APOGEE DR16 and one LAMOST DR5) and $f(M)=0.048\ M_{\odot}$. Given the \texttt{StarHorse} mass $M_L=1.07\ M_{\odot}$ this implies a companion mass $M_c\sim0.49\ M_{\odot}$. While this companion is unlikely to be a black hole this system does offer a sanity check on our minimum mass estimation of $M_{c,\rm{min}} = 0.39\ M_{\odot}$. ASASSN-V J190444.42+392718.4 has one radial velocity measurement from APOGEE DR16 and one from LAMOST DR5 that are separated by 0.44 in phase. The mass function is $f(M)=0.33\ M_{\odot}$, suggesting a companion mass $M_c = 1.23\ M_{\odot}$ for the \texttt{StarHorse} luminous star mass $M_*=1.14\ M_{\odot}$. This is again consistent with the minimum mass $M_{c,\rm{min}}= 0.61\ M_{\odot}$ calculated from the ELL analytical model. 

There are \nRVbad{} ELL candidates that have $\rm{Max}(\Delta\phi)>0.25$ that are not well-fit by the circular orbit model: \nameRVbad{}. Two of these, ASASSN-V J080826.59-055109.9 and  J084328.53+402247.5, have LAMOST observations with maximum $\Delta(\rm{RV})$ comparable to the radial velocity uncertainty. J061635.56+231909.4 has three LAMOST points, two of which have $\Delta\phi<0.05$ but $\Delta(\rm{RV})>128$~km/s. The four APOGEE RV measurements for J161803.43+420416.8 have a maximum $\Delta(\rm{RV})=6.2$~km/s and are also not well-fit by circular orbit model. While a poor RV fit may indicate that the system is not an ELL, we do not remove any ELL candidates on the basis of the sparse APOGEE/LAMOST RV measurements. These systems are listed in Table \ref{tab:rv_table} but we do not report the RV fits.

Finally, we cross-match our ELL candidate catalog with the Master X-Ray Catalog provided by HEASARC\footnote{\url{https://heasarc.gsfc.nasa.gov/W3Browse/all/xray.html}} and with the Swift X-ray Telescope Point Source Catalog \citep[2XPs, ][]{Evans20}. We found that \nXray{} of the ELL candidates are likely X-ray sources, although the larger offset ROSAT matches should probably be confirmed. Table \ref{tab:ELLxray} lists the sources and their estimated X-ray luminosities given the \textit{Gaia} EDR3 parallaxes.

X-ray emission could correspond to accretion in a compact object binary. In their quiescent states, X-ray binaries have been detected with relatively low luminosities $<10^{32}$ erg/s. \citet{Rodriguez20} reported the X-ray observations of GS 2000$+$25 in the quiescent state with luminosity $L=1.1^{+1.0}_{-0.7}\times10^{30} (d/2\rm{\ kpc})^2$ erg/s. Alternatively, high-energy emission may be evidence of chromospheric activity or coronal emission in eclipsing binaries \citep[e.g., ][]{Bedford90}. The presence of X-ray emission does not offer clear distinctions between ELL and non-ELL variables, but may serve as a useful metric for prioritizing RV followup of ELL candidates. 

\begin{table*}
\caption{ELL Candidates with X-Ray cross-matches from the Master X-ray Catalog and Swift 2XPS.}
\label{tab:ELLxray}
\begin{tabular}{llrrr}
\toprule
               {ASAS-SN Name} & {Telescope} & {Period (days)} & {X-ray Luminosity (erg/s)} & {Separation (arcsecond)} \\
\midrule
 ASASSN-V J172630.51-381304.4 &  XMM-NEWTON &           18.31 &        $3.37 \times 10^{30}$ &                     1.19 \\
 ASASSN-V J015028.29+363449.2 &       ROSAT &           17.39 &        $1.80 \times 10^{31}$ &                    13.27 \\
 ASASSN-V J010522.46+482502.4 &       Swift &           16.85 &        $2.11 \times 10^{32}$ &                     1.77 \\
 ASASSN-V J161454.69-513604.8 &  XMM-NEWTON &           15.76 &        $3.76 \times 10^{30}$ &                     3.65 \\
 ASASSN-V J155416.74+081826.6 &       Swift &           10.09 &        $2.09 \times 10^{31}$ &                     1.95 \\
 ASASSN-V J032202.62-511248.8 &     CHANDRA &            8.74 &        $5.24 \times 10^{30}$ &                     0.34 \\
 ASASSN-V J183142.73+444829.1 &       ROSAT &            0.41 &        $1.34 \times 10^{31}$ &                     9.30 \\
 ASASSN-V J023618.76+615619.7 &  XMM-NEWTON &            0.31 &        $3.10 \times 10^{30}$ &                     1.29 \\
 ASASSN-V J060831.46-442417.5 &       ROSAT &            0.28 &        $5.54 \times 10^{30}$ &                     4.71 \\
\bottomrule
\end{tabular}

\end{table*}

\section{Conclusions} \label{sec:summary}

Previous searches for non-interacting compact object binaries have started from radial velocity surveys to identify variables with a large mass function \citep{Thompson19, Liu19, Jayasinghe21}. Here we take advantage of multi-year all-sky photometry from ASAS-SN to search almost 200,000 variables for ellipsoidal variability. The final result is a catalog of \nELL{} ELL candidates and \nECL{} ELL+ECL candidates.

Our search method identifies light curves with uneven minima by comparing an analytical model of ellipsoidal modulations to a cosine fit. We perform multiple rounds of visual inspection to remove eclipsing binaries and rotational variables from the catalog. By combining the analytical model with masses and radii from the \citet{Anders19} {\tt StarHorse} catalog, we derive conservative estimate of the minimum companion masses for the ELL candidates. 

The subset of ELL candidates with multiple RV measurements offers a sanity check on the minimum mass as a metric to sort ELL candidates. Where we can make the comparison, the minimum companion mass calculated from the ELL analytical model is consistent with the mass estimated from the RV semi-amplitude, suggesting the minimum companion mass is a viable metric to sort ELL candidates for follow-up. \citet{Jayasinghe21} report that V723~Mon has $M=3.3^{+2.8}_{-0.7}\ M_{\odot}$ and our minimum mass estimate is $M_{c,\rm{min}} = 1.39~M_\odot$. The majority of our ELL candidates have a lower minimum companion mass, suggesting that systems like V723~Mon are rare. Systems with higher companion mass limits are likely the best targets for radial velocity follow-up in the search for non-interacting compact objects. 

\section*{Acknowledgements}

We thank the Las Cumbres Observatory and its staff for its continuing support of the ASAS-SN project. 
The ASAS-SN operations are supported by the Gordon and Betty Moore Foundation through grant GBMF5490 to the Ohio State University.

DMR is supported by The Ohio State University Graduate Fellowship.
KZS, TJ, and CSK are supported by NSF grants AST-1814440 and 
AST-1908570. TJ acknowledges support from the Ohio State Presidential Fellowship. TAT is supported in part by NASA grant 80NSSC20K0531. 
Support for JLP is provided in part by the
Ministry of Economy, Development, and Tourism's Millennium Science 
Initiative through grant IC120009, awarded to The Millennium Institute 
of Astrophysics, MAS. 

This work has made use of data from the European Space Agency (ESA) mission
{\it Gaia} (\url{https://www.cosmos.esa.int/gaia}), processed by the {\it Gaia}
Data Processing and Analysis Consortium (DPAC,
\url{https://www.cosmos.esa.int/web/gaia/dpac/consortium}). Funding for the DPAC
has been provided by national institutions, in particular the institutions
participating in the {\it Gaia} Multilateral Agreement. This research has made use of the SIMBAD database, operated at CDS, Strasbourg, France. Guoshoujing Telescope (the Large Sky Area Multi-Object Fiber Spectroscopic Telescope LAMOST) is a National Major Scientific Project built by the Chinese Academy of Sciences. Funding for the project has been provided by the National Development and Reform Commission. LAMOST is operated and managed by the National Astronomical Observatories, Chinese Academy of Sciences. 

Funding for the Sloan Digital Sky 
Survey IV has been provided by the 
Alfred P. Sloan Foundation, the U.S. 
Department of Energy Office of 
Science, and the Participating 
Institutions. 

SDSS-IV acknowledges support and 
resources from the Center for High 
Performance Computing  at the 
University of Utah. The SDSS 
website is www.sdss.org.

SDSS-IV is managed by the 
Astrophysical Research Consortium 
for the Participating Institutions 
of the SDSS Collaboration including 
the Brazilian Participation Group, 
the Carnegie Institution for Science, 
Carnegie Mellon University, Center for 
Astrophysics | Harvard \& 
Smithsonian, the Chilean Participation 
Group, the French Participation Group, 
Instituto de Astrof\'isica de 
Canarias, The Johns Hopkins 
University, Kavli Institute for the 
Physics and Mathematics of the 
Universe (IPMU) / University of 
Tokyo, the Korean Participation Group, 
Lawrence Berkeley National Laboratory, 
Leibniz Institut f\"ur Astrophysik 
Potsdam (AIP),  Max-Planck-Institut 
f\"ur Astronomie (MPIA Heidelberg), 
Max-Planck-Institut f\"ur 
Astrophysik (MPA Garching), 
Max-Planck-Institut f\"ur 
Extraterrestrische Physik (MPE), 
National Astronomical Observatories of 
China, New Mexico State University, 
New York University, University of 
Notre Dame, Observat\'ario 
Nacional / MCTI, The Ohio State 
University, Pennsylvania State 
University, Shanghai 
Astronomical Observatory, United 
Kingdom Participation Group, 
Universidad Nacional Aut\'onoma 
de M\'exico, University of Arizona, 
University of Colorado Boulder, 
University of Oxford, University of 
Portsmouth, University of Utah, 
University of Virginia, University 
of Washington, University of 
Wisconsin, Vanderbilt University, 
and Yale University. This research made use of Astropy, a community-developed core Python package for Astronomy. 

\section*{Data Availability}

The ASAS-SN photometric data underlying this article are available on the ASAS-SN Photometry Data base (https://asas-sn.os
u.edu/photometry) and the ASAS-SN variable stars data base
(https://asas-sn.osu.edu/variables). The spectroscopic
data sets underlying this article were accessed from sources in the
public domain: APOGEE (https://www.sdss.org/dr16/), LAMOST
(http://dr5.lamost.org/), and RAVE (https://www.rave-survey.org/). The data underlying this article are available in the article and in its online supplementary material.


\clearpage
\bibliographystyle{mnras}
\bibliography{ell} 






\bsp	
\label{lastpage}
\end{document}